\documentclass{article}
\usepackage{amsmath}
\usepackage{amsfonts}
\usepackage{amsthm}
\usepackage{amssymb}
\usepackage{amscd}
\usepackage{color}
\usepackage{a4wide}
\usepackage{bm}
\usepackage{latexsym}
\usepackage{indentfirst}
\usepackage[latin1]{inputenc}
\usepackage{ulem}

\def\qed{\hfill $\blacksquare$}

\newtheorem{definition}{\underline{Definition}}[section]
\newtheorem{proposition}[definition]{Proposition}
\newtheorem{theorem}[definition]{Theorem}
\newtheorem{lema}[definition]{Lemma}

\newtheorem{assumption}[definition]{Assumption}

\theoremstyle{definition}

\numberwithin{equation}{section}

\begin{document}
\begin{center}
{\LARGE \bf  A rigorous proof of the Landau-Peierls formula and much more}

\medskip

\today
\end{center}

\begin{center}
\small{
Philippe Briet\footnote{Universit{\'e} du sud Toulon-Var \& Centre de Physique Th\'eorique, Campus de Luminy, Case 907
13288 Marseille cedex 9, France; e-mail: briet@univ-tln.fr  },
Horia D. Cornean\footnote{Department of Mathematical Sciences,
    Aalborg
    University, Fredrik Bajers Vej 7G, 9220 Aalborg, Denmark; e-mail:
    cornean@math.aau.dk},
Baptiste Savoie\footnote{Centre de Physique Th\'eorique,
Campus de Luminy, Case 907
13288 Marseille cedex 9, France, and Department of Mathematical Sciences,
    Aalborg
    University, Fredrik Bajers Vej 7G, 9220 Aalborg, Denmark; e-mail:
    baptiste.savoie@gmail.com}}

\end{center}

\vspace{0.5cm}

\begin{abstract}
We present a rigorous mathematical treatment of the zero-field
orbital magnetic susceptibility of a non-interacting Bloch
electron gas,
at fixed temperature and density, for both metals and
semiconductors/insulators. In particular,
we obtain the Landau-Peierls formula in
the low temperature and density limit as conjectured by T. Kjeldaas and
W. Kohn in 1957.

\end{abstract}

\vspace{0.5cm}

\tableofcontents

\section{Introduction and the main results}
Understanding the zero-field magnetic susceptibility of a Bloch electron gas
is one of the oldest problems in quantum
statistical mechanics.

The story began in 1930 with a paper by
L. Landau \cite{Lan}, in which he computed the diamagnetic
susceptibility of a free degenerate gas. (Note that the rigorous
proof of Landau's formula for free electrons
was given by Angelescu {\it et al.} \cite{ABN} and came as late as 1975.)
For Bloch electrons (which are subjected to a periodic background
electric potential), the problem is much harder and -to our best
knowledge- it has not been solved yet in its full generality.

The first important contribution to the periodic problem came in 1933,
when R. Peierls
\cite{Pei} introduced his celebrated Peierls substitution and
constructed an effective band Hamiltonian which permitted to reduce
the problem to free electrons. Needless to say that
working with only one energy band instead of the full magnetic Schr\"odinger
operator is an important simplification. Under the
tight-binding approximation he claimed that the dominant
contribution to the zero-field orbital susceptibility of a Bloch electron gas
in metals (at zero temperature) is purely diamagnetic and is given
by the so-called Landau-Peierls formula which consists of replacing
in the Landau formula the mass by the effective mass of the
electron. He showed as well the existence of another contribution
which has no simple interpretation and whose magnitude and sign are uncertain.

In 1953, E.N. Adams \cite{Adams} claimed that the
Landau-Peierls susceptibility is not always the dominant contribution
to the zero-field orbital susceptibility. By considering the case of 'simple
metals' (for which the tight-binding approximation is not appropriate), he showed that there exists others contributions (certain have even positive sign!) coming from the bands not containing the Fermi energy. Besides in special cases these contributions are of the same order of magnitude than the Landau-Peierls formula. However any general formulation of these contributions were stated.

In 1957, T. Kjeldaas and W. Kohn \cite{KK} were probably the first ones who
suggested that for 'simple metals' {\it the Landau-Peierls approximation is only valid
in the limit of weak density of electrons}, moreover, the Landau-Peierls formula
(see below \eqref{sulp} and \eqref{chilp}) has to be corrected with some
higher order terms in the particle density, and these terms must come
from the bands not containing the Fermi energy.

These three papers generated a lot of activity,
where the goal was to write down an exact expression for the
  zero-field magnetic susceptibility of a Bloch electron gas in metals
  at zero temperature. In what follows we comment on some of the most
  important works.

The first attempt to address the full quantum mechanical problem -even
though the carriers were boltzons and not fermions- was made by
J.E. Hebborn and E.H. Sondheimer \cite{HS1, HS2} in 1960. Unlike the previous
authors, they developed a magnetic perturbation theory for the
trace per unit volume  defining the pressure. The biggest problem of
their formalism is that they assumed that all Bloch energy bands are
not overlapping (this is generically false; for a proof of the
Bethe-Sommerfeld conjecture in dimension $3$ see e.g. \cite{HeMo}),
and that the Bloch basis is smooth in the quasi-momentum variables. This assumption can
fail at the points where the energy bands cross each other. Not to
mention that no convergence issues were addressed in any way.

In 1962, L.M. Roth \cite{Rot} developed a sort of magnetic
pseudodifferential calculus starting from the ideas of Peierls,
Kjeldaas and Kohn. She used this formalism in order to compute
local traces and magnetic expansions. Similar results are obtain by
E.I. Blunt \cite{BL}. Their formal computations can most probably be made
rigorous in the case of simple bands.

In 1964, Hebborn {\it et al.} \cite{HLS} simplified the formalism
developed in \cite{HS2} and gave for the first time a formula for the
zero-field susceptibility of a boltzon gas. Even though the proofs
lack any formal rigor, we believe that their derivation could be made
rigorous for systems where the Bloch bands do not overlap. But this is
generically not the case.

The same year, G.H. Wannier and U.N. Upadhyaya \cite{WU} go back to the method
advocated by Peierls, and replace the true magnetic Schr\"odinger
operator with a (possibly infinite) number of bands modified with the
Peierls phase factor. They claim that their result is equivalent
with that one of Hebborn and Sondheimer \cite{HS2}, but no details are
given. Anyhow, the result uses in an essential way the non-overlapping
of Bloch bands. At the same time, L. Glasser
  \cite{Gla} gave an expression of the bulk zero-field susceptibility
  in terms of effective mass by the usual nearly free electron approximation.

In 1969, P.K. Misra and L.M. Roth \cite{MR} combined the method of \cite{Rot}
with the ideas of Wannier in order to include the core electrons in
the computation.

In 1972, P.K. Misra and L. Kleinman \cite{MK} had the very nice idea of using
sum-rules in order to replace derivatives with respect to the
quasi-momentum variables, with matrix elements of the "true" momentum
operator. They manage in this way to rewrite the formulas previously derived by
Misra and Roth (which only made sense for non-overlapping bands) in a form which
might also hold for overlapping bands.

As we have already mentioned, the first serious mathematical approach
on the zero-field susceptibility appeared as late as 1975, due to
Angelescu {\it et al.} \cite{ABN}. Then in 1990, B. Helffer and
J. Sj\"ostrand \cite{HeSj} developed for the first time a rigorous theory based on the Peierls
substitution and considered the connection with the de Haas-Van Alphen
effect. These and many more results were reviewed by G. Nenciu in 1991 \cite{N}.
A related  problem in which the electron gas is confined by a trapping potential
was considered by M. Combescure and D. Robert in 2001 \cite{CR}. They obtained the Landau
formula in the limit $\hbar\to 0$.

Finally we mention that the magnetic response can be
described using the semiclassical theory of the orbital magnetism and the Berry-phase formula,
see \cite{Re} for further details. The link between
this approach and our work has yet to be  clarified.

Our current paper is based on what we call magnetic perturbation
theory, as developed by the authors and their collaborators
in a series of papers starting with 2000 (see \cite{CN0, Cor1, CNP, CN3, CN,
  BC, BrCoLo1, BrCoLo2, BrCoLo3, BCS1}
and references therein). The results we obtain in Theorem
\ref{maintheorem} give a complete
answer to the problem of zero-field susceptibility. Let us now discuss
the setting and properly formulate the mathematical problem.

\subsection{The setting}

Consider a confined quantum gas of charged particles
obeying the Fermi-Dirac statistics. The spin is not considered since
we are only interested in orbital magnetism. Assume that the gas is subjected
to a constant magnetic field and an external periodic electric
potential. The interactions between particles are neglected and the
gas is at thermal equilibrium.

The gas is trapped in a large cubic box, which is given by
$\Lambda_{L} = \big(-\frac{L}{2},\frac{L}{2}\big)^{3}$, $L \geq
1$.

Let us introduce our one-body Hamiltonian. We consider a uniform
magnetic field $\mathbf{B} = (0,0,B)$ with $B \geq 0$, parallel to the
third direction of the canonical basis of $\mathbb{R}^{3}$. Let
$\mathbf{a}(\mathbf{x})$ be the symmetric (transverse) gauge
$\mathbf{a}(\mathbf{x}) :=\frac{1}{2}(-x_{2},x_{1},0)$ which generates the magnetic field $(0,0,1)$.

We consider that the background electric potential $V$ is smooth, i.e.
$V \in \mathcal{C}^{\infty}(\mathbb{R}^{3})$ is a real-valued function
and periodic with respect to a (Bravais) lattice $\Upsilon$ with unit
cell $\Omega$. Without loss of generality, we assume that $\Upsilon$ is the
cubic lattice $\mathbb{Z}^{3}$, thus $\Omega$ is the unit cube centered
at the origin of coordinates.

When the box is finite i.e. $1 \leq L < \infty$, the dynamics of each
particle is determined by a Hamiltonian defined in $L^{2}(\Lambda_{L})$
with Dirichlet boundary conditions on $\partial\Lambda_{L}$:
\begin{equation}
\label{hL}
H_{L}(\omega) = \frac{1}{2} \big(-i\nabla_{\mathbf{x}} - \omega\mathbf{a}(\mathbf{x})\big)^{2} + V_{L}(\mathbf{x})
\end{equation}
where $V_{L}$ stands for the restriction of $V$ to the box
$\Lambda_{L}$. Here $\omega := \frac{e}{c}B \in
\mathbb{R}$ denotes the cyclotron frequency.
The operator $H_{L}(\omega)$ is self-adjoint on the domain
$D(H_{L}(\omega)\big) = \mathcal{H}_{0}^{1}(\Lambda_{L}) \cap \mathcal{H}^{2}(\Lambda_{L})$.
It is well known (see \cite{RS2}) that $H_{L}(\omega)$ is bounded
from below and has compact resolvent. This implies that its spectrum
is purely discrete with an accumulation point at infinity. We denote
the set of eigenvalues (counting multiplicities and in increasing
order) by $\{e_{j}(\omega)\}_{j \geq 1}$.

When $L=\infty$ we denote by $H_{\infty}(\omega)$ the
unique self-adjoint extension of the operator
\begin{equation}
\label{hinfini}
\frac{1}{2}\big(-i\nabla_{\mathbf{x}} - \omega\mathbf{a}(\mathbf{x})\big)^{2} + V(\mathbf{x})
\end{equation}
initially defined on $\mathcal{C}_{0}^{\infty}(\mathbb{R}^{3})$. Then
$H_{\infty}(\omega)$ is bounded from below and only
has essential spectrum (see e.g. \cite{BCZ}).

Now let us define some quantum statistical quantities related to the quantum gas
introduced above. For the moment we use the grand canonical
formalism. The finite volume pressure and density of our quantum gas
at inverse temperature $\beta := (k_{B}T)^{-1} > 0$ ($k_{B}$ stands
for the Boltzmann constant), at fugacity $z:= \mathrm{e}^{\beta \mu} >
0$ ($\mu \in \mathbb{R}$ stands for the chemical potential) and at
cyclotron frequency $\omega \in \mathbb{R}$ are given by (see e.g. \cite{H}):
\begin{align}
\label{pressurefv}
P_{L}(\beta,z,\omega) &:= \frac{1}{\beta \vert \Lambda_{L}\vert} \mathrm{Tr}_{L^{2}(\Lambda_{L})} \big\{\ln\big(\mathbf{1} + z \mathrm{e}^{-\beta H_{L}(\omega)}\big)\big\} = \frac{1}{\beta \vert \Lambda_{L}\vert} \sum_{j=1}^{\infty} \ln\big(1 + z \mathrm{e}^{-\beta e_{j}(\omega)}\big)\\
\label{densityfv}
\rho_{L}(\beta,z,\omega) &:= \beta z \frac{\partial P_{L}}{\partial
  z}(\beta,z,\omega) = \frac{1}{\vert \Lambda_{L}\vert}
\sum_{j=1}^{\infty} \frac{z \mathrm{e}^{-\beta e_{j}(\omega)}}{1 + z\mathrm{e}^{-\beta e_{j}(\omega)}}.
\end{align}
As the semi-group $\mathrm{e}^{-\beta H_{L}(\omega)}$ is trace class,
the series in \eqref{pressurefv} and \eqref{densityfv}
are absolutely convergent. Since the function $\mathbb{R} \owns \omega
\mapsto P_{L}(\beta,z,\omega)$ is smooth (see \cite{BCS1}), we can define the
finite volume orbital susceptibility as the second derivative of
the pressure with respect to the intensity $B$ of the magnetic field at $B=0$
(see e.g. \cite{ABN}):
\begin{equation}
\label{susceptifv}
\mathcal{X}_{L}^{GC}(\beta,z) := \bigg(\frac{e}{c}\bigg)^{2} \frac{\partial^{2} P_{L}}{\partial \omega^{2}}(\beta,z,0).
\end{equation}
When $\Lambda_{L}$ fills the whole space, we proved in \cite{BCS2}
that the thermodynamic limits of the three grand
canonical quantities defined above exist. By denoting
$P_{\infty}(\beta,z,\omega) := \lim_{L \rightarrow \infty}
P_{L}(\beta,z,\omega)$, we proved moreover
the following pointwise convergence:
\begin{align}
\label{densitylim}
\rho_{\infty}(\beta,z,\omega) &:= \beta z \frac{\partial P_{\infty}}{\partial z}(\beta,z,\omega) = \lim_{L \rightarrow \infty} \beta z  \frac{\partial P_{L}}{\partial z}(\beta,z,\omega) \\
\label{susceptilim}
\mathcal{X}_{\infty}^{GC}(\beta,z) &:= \bigg(\frac{e}{c}\bigg)^{2} \frac{\partial^{2} P_{\infty}}{\partial \omega^{2}}(\beta,z,0) = \lim_{L \rightarrow \infty} \bigg(\frac{e}{c}\bigg)^{2} \frac{\partial^{2} P_{L}}{\partial \omega^{2}}(\beta,z,0)
\end{align}
and the limit commutes with the first derivative (resp. the second derivative) of the grand canonical pressure with respect to the fugacity $z$ (resp. to the external magnetic field $B$).

Now assume that our fixed
external parameter is the density of particles $\rho_{0} > 0$.
We prefer to see $\rho_{\infty}$ as a
function of the chemical potential $\mu$ instead of the fugacity $z$;
the density is a strictly increasing function with respect to both $\mu$ and $z$. Denote by
$\mu_{\infty}(\beta,\rho_{0}) \in \mathbb{R}$ the unique solution of the equation:
\begin{equation}
\label{muinfini}
\rho_{0}=\rho_{\infty} \left (\beta,e^{\beta\mu_{\infty}(\beta,\rho_{0})},
  0\right ).
\end{equation}
The bulk orbital susceptibility at $\beta >
0$ and fixed density $\rho_{0} > 0$ defined from \eqref{susceptilim}
is defined as:
\begin{equation}
\label{susceptic}
\mathcal{X}(\beta,\rho_{0}) := \mathcal{X}_{\infty}^{GC}\left
  (\beta,e^{\beta \mu_{\infty}(\beta,\rho_{0})}\right ).
\end{equation}
In fact one can also show that
$\mathcal{X}(\beta,\rho_{0}) =
-\big(\frac{e}{c}\big)^{2} \frac{\partial^{2} f_{\infty}}{\partial \omega^{2}}(\beta,\rho_{0},0)$
where $f_{\infty}(\beta,\rho_{0},\omega)$ is the thermodynamic limit
of the reduced free energy defined as the Legendre transform of the
thermodynamic limit of the pressure (see e.g.\cite{Ru}).
Note that for a perfect quantum gas and in the limit
  of low temperatures, \eqref{susceptic}
leads to the so-called Landau diamagnetic susceptibility, see e.g. \cite{ABN}.

In order to formulate our main result, we need to introduce some more
notation. In the case in which $\omega=0$, the Floquet theory for
periodic operators (see e.g. \cite{BS}, \cite{Ku} and section 3) allows one to use
the band structure of the spectrum of $H_{\infty}(0)$. Denote by
$\Omega^{*} = 2\pi \Omega$ the Brillouin zone of the dual
lattice
$\Upsilon^{*} \equiv 2\pi \mathbb{Z}^{3}$.

If $j \geq 1$,
the $j$th Bloch band function is defined by $\mathcal{E}_{j} :=
[\min_{\mathbf{k} \in \Omega^{*}}E_{j}(\mathbf{k}),\max_{\mathbf{k} \in
  \Omega^{*}} E_{j}(\mathbf{k})]$ where $\{E_{j}(\mathbf{k})\}_{j \geq 1}$
is the set of eigenvalues (counting multiplicities and in {\it increasing}
order) of the fiber Hamiltonian $h(\mathbf{k}) := \frac{1}{2}(-i\nabla
+\mathbf{k})^{2} + V$ living in $L^{2}(\mathbb{T}^{3})$ with
$\mathbb{T}^{3} := \mathbb{R}^{3}/\mathbb{Z}^{3}$ the 3-dimensional
torus. With this definition, the Bloch energies $E_{j}(\cdot)$ are
continuous on the whole of $\Omega^{*}$, but they are differentiable
only outside a zero Lebesgue measure subset of $\Omega^{*}$
corresponding to cross-points. In the following we make the assumption
that the $E_j$'s are simple eigenvalues
for $\mathbf{k}$ in a subset of $\Omega^{*}$ with full measure. Note
that this assumption is not essential for our approach but it
simplifies the presentation, see Remark 3 below the Theorem \ref{maintheorem}.

The spectrum of $H_{\infty}(0)$
is absolutely continuous and given (as a set of points) by
$\sigma(H_{\infty}(0)) = \bigcup_{j=1}^{\infty} \mathcal{E}_{j}$.
Note that the sets $\mathcal{E}_{j}$ can overlap each other in many ways, and some
of them can even coincide even though they are images of increasingly ordered
functions. The energy bands are disjoint unions of
$\mathcal{E}_{j}$'s. Moreover, if $\max \mathcal{E}_{j} < \min
\mathcal{E}_{j+1} $ for some $j\geq 1$ then we have a spectral
gap. Since the Bethe-Sommerfeld conjecture holds true
  under our conditions \cite{HeMo},
the number of spectral gaps is finite, if not zero.

It remains to introduce the integrated density of states of the
operator $H_{\infty}(0)$. Recall its definition. For any $E \in
\mathbb{R}$, let $N_{L}(E)$ be the number of eigenvalues of
$H_{{L}}(0)$ not greater than $E$. The integrated density of
states of $H_{\infty}(0)$ is defined by the limit (see \cite{DIM}):
\begin{equation}
\label{ids}
n_{\infty}(E) := \lim_{L \rightarrow  \infty} \frac{N_{L}(E)}{\vert \Lambda_{L} \vert} = \lim_{L \rightarrow  \infty} \frac{\mathrm{Tr}\big\{\chi_{(-\infty,E]}(H_{L}(0))\big\}}{\vert \Lambda_{L} \vert}
\end{equation}
and $n_{\infty}(\cdot)$ is a positive continuous and non-decreasing
function (see e.g. \cite{BS}). In this case one can express
$n_\infty(E)$ with the help of the Bloch energies in the following
way:
\begin{equation}
\label{ids'}
n_{\infty}(E)=\frac{1}{(2\pi)^3}\sum_{j\geq
  1}\int_{\Omega^*}\chi_{[E_{0}, E]}(E_{j}(\mathbf{k}))\; d{\bf k}
\end{equation}
where $\chi_{[E_{0},E]}(\cdot)$ is the characteristic function of the interval $[E_{0},E]$. Thus
$n_{\infty}$ is clearly continuous in $E$ due to the continuity of the Bloch
bands. Moreover, this function is piecewise constant when $E$ belongs
to a spectral gap.

\subsection{The statements of our main results}

The first theorem is not directly related to the magnetic problem,
and it deals with the rigorous definition of the Fermi energy for
Bloch electrons. Even though these results are part of the 'physics folklore',
we have not found a serious mathematical treatment in the literature.

\begin{theorem}\label{metalsisemi}
 Let $\rho_{0} > 0$ be fixed. If $\mu_{\infty}(\beta,\rho_{0})$
is the unique real solution of the
equation $\rho_{\infty}\left
  (\beta,e^{\beta\mu},0\right ) = \rho_{0}$ (see \eqref{muinfini}),
then the limit:
\begin{equation}
\label{ferenergy}
\mathcal{E}_{F}(\rho_{0}) :=
\lim_{\beta \rightarrow \infty} \mu_{\infty}(\beta,\rho_{0})
\end{equation}
exists and defines an
increasing function of $\rho_{0}$ called the Fermi energy. There can only occur two
cases:

{\bf SC} (semiconductor/insulator/semimetal): Suppose that there exists some
$N \in \mathbb{N}^{*}$ such that $\rho_{0} =
n_{\infty}(E)$ for all $E\in [\max \mathcal{E}_{N}, \min
\mathcal{E}_{N+1}]$. Then:
\begin{equation}
\label{gapfermi}
\mathcal{E}_{F}(\rho_{0}) = \frac{\max \mathcal{E}_{N} +
\min \mathcal{E}_{N+1}}{2}.
\end{equation}

{\bf M} (metal): Suppose that there exists a unique solution $E_M$ of the equation $
n_{\infty}(E_M)=\rho_{0}$ which belongs to $(\min \mathcal{E}_{N}, \max \mathcal{E}_{N})$
for some (possibly not unique) $N$. Then :
\begin{equation}
\label{metalfermi}
\mathcal{E}_{F}(\rho_{0}) = E_M.
\end{equation}
\end{theorem}

\vspace{0.5cm}

\noindent\textbf{\textit{Remark 1}}. In other words, a semiconductor/semimetal
either has its Fermi energy in the middle of
a non-trivial gap (this occurs if $\max \mathcal{E}_{N} < \min
\mathcal{E}_{N+1}$), or where the two consecutive Bloch bands touch each
other closing the gap (this occurs if $\max \mathcal{E}_{N} = \min
\mathcal{E}_{N+1}$). As for a metal, its Fermi energy lies
in the interior of a Bloch
band.

\noindent\textbf{\textit{Remark 2}}. According to the above result,
$\mathcal{E}_{F}$ is discontinuous at all values of $\rho_{0}$ for which the equation
$n_{\infty}(E)=\rho_{0}$ does not have a unique solution. Each open gap gives such a
discontinuity.

\vspace{0.5cm}

Now here is our main result concerning the orbital susceptibility of a Bloch
electrons gas at fixed density and zero temperature:

\begin{theorem}
\label{maintheorem}
Denote by $E_{0} := \inf \sigma(H_{\infty}(0))$.

\indent {\rm (i)}. Assume that the Fermi energy is in the middle of a
non-trivial gap (see \eqref{gapfermi}). Then there exist $2N$
functions $\mathfrak{c}_{j}(\cdot\,), \mathfrak{d}_{j}(\cdot\,)$, with $1\leq j \leq N$, defined on $\Omega^{*}$
outside a set of Lebesgue measure zero, such that the integrand in
\eqref{susemicon} can be
extended by continuity to the whole of $\Omega^{*}$ and:
\begin{align}
\label{susemicon}
\mathcal{X}_{\rm SC}(\rho_{0}):=\lim_{\beta \rightarrow \infty} \mathcal{X}(\beta,\rho_{0}) =
\bigg(\frac{e}{c}\bigg)^{2} \frac{1}{2}  \frac{1}
{(2\pi)^3} \int_{\Omega^{*}} \mathrm{d}\mathbf{k}\, \sum_{j=1}^{N} \bigg\{\mathfrak{c}_{j}(\mathbf{k}) + \big\{E_{j}(\mathbf{k}) - \mathcal{E}_{F}(\rho_{0})\big\}\mathfrak{d}_{j}(\mathbf{k})\bigg\}.
\end{align}
\indent {\rm (ii)}. Suppose that there exists a unique $N\geq 1$
such that $\mathcal{E}_{F}(\rho_{0})\in
(\min \mathcal{E}_N, \max \mathcal{E}_N)$.
Assume that the Fermi surface
$\mathcal{S}_{F}:=\{\mathbf{k} \in \Omega^{*}\,:\, E_{N}(\mathbf{k}) =
\mathcal{E}_{F}(\rho_{0})\}$ is smooth and non-degenerate. Then there
exist $2N+1$ functions
$\mathcal{F}_{N}(\cdot),\mathfrak{c}_{j}(\cdot),\mathfrak{d}_{j}(\cdot)$ with
$1 \leq j \leq N$, defined on
$\Omega^*$ outside a set of Lebesgue measure zero, in such a way that they are
all continuous on $\mathcal{S}_{F}$ while the second integrand in
\eqref{sumetal} can be
extended by continuity to the whole of $\Omega^*$:
\begin{align}
\label{sumetal}
&\mathcal{X}_{\rm M}(\rho_{0}):=\lim_{\beta \rightarrow \infty} \mathcal{X}(\beta,\rho_{0}) = - \bigg(\frac{e}{c}\bigg)^{2} \frac{1}{12} \frac{1}{(2\pi)^3} \\
& \bigg\{\int_{\mathcal{S}_{F}}
\frac{\mathrm{d}\sigma(\mathbf{k})}{\big\vert \nabla E_{N}(\mathbf{k})
  \big\vert}\,\bigg[\frac{\partial^{2}E_{N}(\mathbf{k})}{\partial
  k_{1}^{2}}\; \frac{\partial^{2}E_{N}(\mathbf{k})}{\partial
  k_{2}^{2}} - \bigg(\frac{\partial^{2}E_{N}(\mathbf{k})}{\partial
  k_{1}\partial k_{2}}\bigg)^{2} - 3 \mathcal{F}_{N}(\mathbf{k})\bigg]
\nonumber \\
&- 6 \int_{\Omega^{*}} \mathrm{d}\mathbf{k}\, \sum_{j=1}^{N} \bigg[
\chi_{[E_{0},\mathcal{E}_{F}(\rho_{0})]}\big(E_{j}(\mathbf{k})\big)
\mathfrak{c}_{j}(\mathbf{k}) + \big\{E_{j}(\mathbf{k}) -
\mathcal{E}_{F}(\rho_{0})\big\}
\chi_{[E_{0},\mathcal{E}_{F}(\rho_{0})]}\big(E_{j}(\mathbf{k})\big)
\mathfrak{d}_{j}(\mathbf{k})\bigg]\bigg\}.\nonumber
\end{align}
Here
$\chi_{[E_{0},\mathcal{E}_{F}(\rho_{0})]}(\cdot\,)$ denotes the
characteristic function of the interval $E_{0}\leq t \leq
\mathcal{E}_{F}(\rho_{0})$.

\indent {\rm (iii)}. Let $k_{F} := (6\pi^{2}
\rho_{0})^{\frac{1}{3}}$ be the Fermi wave vector. Then in the limit
of small densities,
\eqref{sumetal} gives the Landau-Peierls formula:
\begin{equation}
\label{sulp}
\mathcal{X}_{\rm M}(\rho_{0})= - \frac{e^{2}}{24 \pi^{2} c^{2}}\frac{(m_{1}^{*}m_{2}^{*}m_{3}^{*})^{\frac{1}{3}}}{m_{1}^{*} m_{2}^{*}} k_{F} + o(k_{F});
\end{equation}
here $\big[\frac{1}{m_{i}^{*}}\big]_{1 \leq i \leq 3}$ are the
eigenvalues of the positive definite Hessian matrix
$\{\partial_{ij}^2E_1({\bf 0})\}_{1\leq i,j\leq 3}$.
\end{theorem}

\vspace{1cm}

\noindent\textbf{\textit{Remark 1}}. The functions $\mathfrak{c}_{j}(\cdot)$
and $\mathfrak{d}_{j}(\cdot)$ with $1 \leq j \leq N$ which appear in
\eqref{susemicon} are the same as the ones in \eqref{sumetal}.
All of them (as well as $\mathcal{F}_{N}(\cdot)$)
can be explicitely written down in terms of Bloch energy
functions and their associated eigenfunctions. One can notice in
\eqref{sumetal} the appearance of an explicit term associated with the
$N$th Bloch energy function; it is only this term which will generate the
linear $k_F$ behavior in the Landau-Peierls formula.

\noindent\textbf{\textit{Remark 2}}. The
functions $\mathfrak{c}_{j}(\cdot)$
and $\mathfrak{d}_{j}(\cdot)$ might have local singularities at a
set of Lebesgue measure zero where the Bloch bands might touch each other. But
their combinations entering the integrands above are always
bounded because the individual singularities get
canceled by the sum.

\noindent\textbf{\textit{Remark 3}}.
The results in {\rm (i)} and {\rm (ii)} hold true even if some Bloch
bands are degenerate on a subset of full Lebesgue measure of
$\Omega^*$. But in this case the functions $\mathfrak{c}_{j}(\cdot)$,
$\mathfrak{d}_{j}(\cdot)$ and $\mathcal{F}_{N}(\cdot\,)$ cannot be
expressed in the same way as mentioned in Remark 1. Their expressions
are more complicated and require the use of the orthogonal projection corresponding
to $E_{j}(\cdot)$, see the proof of Lemma \ref{regulcof} for further details.

\noindent\textbf{\textit{Remark 4}}. When $m_{1}^{*} = m_{2}^{*} = m_{3}^{*}=
m^{*}$ holds in {\rm (iii)}, \eqref{sulp} is nothing but the usual
Landau-Peierls susceptibility formula:
\begin{equation}
\label{chilp}
\mathcal{X}_{\rm M}(\rho_{0})\sim - \frac{e^{2}}{24 \pi^{2}m^{*} c^{2}} k_{F} \quad
\textrm{when} \quad k_{F} \rightarrow 0.
\end{equation}
Note that our expression is twice smaller than the one in \cite{Pei} since we do not take into account the degeneracy related to the spin of the Bloch electrons.

\noindent\textbf{\textit{Remark 5}}. The assumption $V \in
\mathcal{C}^{\infty}(\mathbb{T}^{3})$ can be relaxed to $V \in
\mathcal{C}^{r}(\mathbb{T}^{3})$ with
$r \geq 23$. The smoothness of $V$ plays an important role in the absolute
convergence of the series defining $\mathcal{X}(\beta,\rho_{0})$ in Theorem
\ref{genexpand},
before the zero-temperature limit; see \cite{CN}
for a detailed discussion on sum rules and local traces for periodic operators.

\noindent\textbf{\textit{Remark 6}}. The role of magnetic perturbation theory
(see Section 3) is {\it crucial} when one wants to write down a formula for
$\mathcal{X}(\beta,\rho_{0})$ which contains no derivatives with
respect to the quasi-momentum ${\bf k}$. Remember that the Bloch energies
ordered in increasing order
and their corresponding eigenfunctions are not necessarily differentiable at
crossing points.

\noindent\textbf{\textit{Remark 7}}. We do not treat the semi-metal case, in
which the Fermi energy equals $\mathcal{E}_F(\rho_0)= \max
\mathcal{E}_N=\min \mathcal{E}_{N+1}$ for some $N\geq 1$ (see
\eqref{gapfermi}). This remains as a challenging open problem.

\subsection{The content of the paper}
Let us briefly discuss the content of the rest of this paper:
\begin{itemize}
\item In Section 2 we thoroughly analyze the behavior of the chemical potential
$\mu_\infty$ when the temperature goes to zero defining the Fermi energy. These
results are important for our main theorem.

\item In Section 3 we give the most important technical result. Applying
the magnetic perturbation theory we arrive at a
general formula for $\mathcal{X}(\beta,\rho_{0})$ which contains no
derivatives with respect to ${\bf k}$. The strategy is somehow
similar to the one used
in \cite{CNP} for the Faraday effect.
\item In Section 4 we perform the zero temperature limit and separately
analyze the situations in which the Fermi energy is either in an open spectral
gap or inside the spectrum. It contains the proofs of Theorem
\ref{maintheorem} $(i)$ and $(ii)$.

\item In Section 5 we obtain the Landau-Peierls formula by taking the low
density limit. It contains the proof of Theorem
\ref{maintheorem} $(iii)$.

\end{itemize}
\section{The Fermi energy}

This section, which can be read independently of the rest of the paper, is
only concerned with the location of the Fermi energy when the intensity of the magnetic field is zero (i.e. $\omega=0$). In particular, we prove Theorem \ref{metalsisemi}.
Although we assumed in the introduction that $V \in \mathcal{C}^{\infty}(\mathbb{T}^{3})$,
all results of this section can be extended to the case $V \in L^{\infty}(\mathbb{T}^{3})$.

\subsection{Some preparatory results}
Let $\xi \mapsto \mathfrak{f}(\beta,\mu;\xi) := \ln\big(1+\mathrm{e}^{\beta(\mu- \xi)}\big)$
be a holomorphic function on the domain $\{ \xi \in \mathbb{C}\,:\, \Im \xi \in (- \pi/\beta, \pi/\beta)\}$. Let $\Gamma$ the positively oriented simple contour included in
the above domain defined by:
\begin{equation}
\label{contour}
\Gamma := \Big\{\Re \xi \in [\delta,\infty),\,\Im \xi = \pm \frac{\pi}{2\beta}\Big\} \cup \Big\{\Re \xi = \delta ,\,\Im \xi \in \Big[-\frac{\pi}{2 \beta},\frac{\pi}{2\beta}\Big]\Big\},
\end{equation}
where $\delta$ is any real number smaller than $E_{0} := \inf \sigma(H_{\infty}(0))\leq
\inf \sigma(H_{\infty}(\omega))$. In the following we use $\delta := E_{0} - 1$.

The thermodynamic limit of the grand-canonical density at $\beta > 0$,
$\mu \in \mathbb{R}$ and $\omega\geq 0$ is given by (see e.g. \cite{BCZ}):
\begin{equation}
\label{rhodef}
\rho_{\infty}(\beta,e^{\beta\mu},\omega)
= \frac{1}{\vert \Omega \vert} \frac{i}{2\pi}
\mathrm{Tr}_{L^{2}(\mathbb{R}^{3})}\bigg\{\chi_{\Omega} \int_{\Gamma} \mathrm{d}\xi\, \mathfrak{f}_{FD}(\beta,\mu;\xi) (H_{\infty}(\omega) - \xi)^{-1} \bigg\}
\end{equation}
where $\mathfrak{f}_{FD}(\beta,\mu;\xi) :=
-\beta^{-1} \partial_{\xi} \mathfrak{f}(\beta,\mu;\xi) =
(\mathrm{e}^{\beta (\xi - \mu)} +1)^{-1}$
is the Fermi-Dirac distribution function and $\chi_{\Omega}$ denotes the characteristic function of
$\Omega$. We prove in \cite{BCS2} (even for singular potentials)
that $\rho_{\infty}(\beta,\cdot,\omega)$ can be analytically extended to
the domain $\mathbb{C} \setminus (-\infty, -\mathrm{e}^{\beta E_{0}(\omega)}]$.

Now assume that the intensity of the magnetic field is zero ($\omega=0$).
The following proposition (stated without proof since the result is well known),
allows us to rewrite \eqref{rhodef}
only using the Bloch energy functions $\mathbf{k} \mapsto E_{j}(\mathbf{k})$ of $H_{\infty}(0)$:
\begin{proposition}
\label{rhobf}
Let $\beta > 0$ and $\mu \in \mathbb{R}$.
Denote by $\Omega^{*}$ the first Brillouin zone of the dual lattice $2\pi \mathbb{Z}^{3}$.
Then:
\begin{equation}
\label{rhodef2}
\rho_{\infty}(\beta,e^{\beta\mu},0) = \frac{1}{(2\pi)^3} \sum_{j=1}^{\infty} \int_{\Omega^{*}} \mathrm{d}\mathbf{k}\, \mathfrak{f}_{FD}(\beta,\mu; E_{j}(\mathbf{k})).
\end{equation}
\end{proposition}

\vspace{0.5cm}

Note that another useful way to express the grand-canonical density at zero magnetic field
consists in bringing into play the integrated density of states (IDS) of the operator
$H_{\infty}(0)$ (see \eqref{ids} for its definition):
\begin{equation}
\label{rho3}
\rho_{\infty}(\beta,e^{\beta\mu},0) = - \int_{-\infty}^{\infty} \mathrm{d}\lambda\,\frac{\partial \mathfrak{f}_{FD}}{\partial \lambda}(\beta,\mu;\lambda) n_{\infty}(\lambda).
\end{equation}

When the density of particles $\rho_{0} > 0$ becomes the fixed
parameter, the relation between the fugacity and
density can be inverted. This is possible since for all $\beta > 0$,
the map $\rho_{\infty}(\beta,\cdot ,0)$ is
strictly increasing on $(0,\infty)$ and defines a
$\mathcal{C}^{\infty}$-diffeomorphism of this interval onto itself.
Then there exists an unique $z_{\infty}(\beta,\rho_{0}) \in
(0,\infty)$
and therefore an unique $\mu_{\infty}(\beta,\rho_{0}) \in \mathbb{R}$ satisfying:
\begin{equation}
\label{inverted}
\rho_{\infty}(\beta,e^{\beta \mu_{\infty}(\beta,\rho_{0})},0) = \rho_{0}.
\end{equation}

We now are interested in the zero temperature limit. The following proposition (again
stated without proof) is a well known, straightforward consequence of the continuity of
$n_{\infty}(\cdot)$:
\begin{proposition}
Let $\mu \geq E_{0} := \inf\sigma(H_{\infty}(0))$ be fixed. We have the identity:
\begin{equation}
\label{limitrho}
\lim_{\beta \rightarrow \infty} \rho_{\infty}(\beta,e^{\beta\mu},0) = \frac{1}{(2\pi)^3} \sum_{j=1}^{\infty} \int_{\Omega^{*}} \mathrm{d}\mathbf{k}\, \chi_{[E_{0},\mu]}(E_{j}(\mathbf{k})) = n_{\infty}(\mu),
\end{equation}
where $\chi_{[E_{0},\mu]}(\cdot)$ denotes the characteristic function of the interval $[E_{0},\mu]$.
\end{proposition}

\vspace{0.5cm}

We end this paragraph with another preparatory result concerning the behavior of
$n_\infty$ near the edges of a spectral gap. This result is contained in the following lemma:

\begin{lema}
\label{ninr}
Let $\rho_0 >0$ be fixed. Assume
that there exists $N \geq 1$ such
that $n_{\infty}(E) = \rho_{0}$ for all $E$ satisfying
$\max \mathcal{E}_{N} \leq E \leq \min \mathcal{E}_{N+1}$.
We set $a_{N} := \max \mathcal{E}_{N}$ and $b_{N} := \min
\mathcal{E}_{N+1}$. Assume that the gap is open, i.e. $a_N<b_N$.
Then for $\delta>0$ sufficiently small, there exists
a constant $C = C_\delta>0$ such that:
\begin{equation}
\label{idsinf}
n_{\infty}(a_{N}) - n_{\infty}(\lambda) \geq C (a_{N} -
\lambda)^{3} \; {\rm whenever}\; \lambda \in [a_{N} - \delta,a_{N}]
\end{equation}
and
\begin{equation}
\label{idsinf22}
 n_{\infty}(\lambda) -n_{\infty}(b_{N})\geq C(
\lambda-b_N)^{3} \; {\rm whenever}\; \lambda \in [b_{N},b_{N}+\delta].
\end{equation}
\end{lema}
\noindent \textbf{\textit{Proof}}. We only prove \eqref{idsinf}, since the proof of the other
inequality \eqref{idsinf22} is similar. Since $a_N=\max_{{\bf k}\in \Omega^*}E_N({\bf
  k})$, the maximum is attained in a (possibly not unique) point ${\bf
  k}_0$, i.e. $a_N=E_N({\bf k}_0)$. This means that $a_{N}$ is a discrete eigenvalue of
finite multiplicity $1 \leq M \leq N$ of the fiber operator $h({\bf
  k}_0)=\frac{1}{2}(-i\nabla +{\bf k}_0)^2+V$. In particular, $a_N$ is isolated from the rest of the spectrum since we assumed that $a_N<b_N\leq E_{N+1}({\bf k}_0)$. Now when ${\bf k}$
slightly varies around ${\bf k}_0$, the eigenvalue $a_N$ will split
into at most $M$ different eigenvalues, the largest of which being
$E_N({\bf k})$. Thus from the second equality in \eqref{limitrho} we obtain:
$$n_{\infty}(a_{N}) - n_{\infty}(\lambda)\geq \frac{1}{(2\pi)^3}{\rm Vol}\{{\bf k}\in
\Omega^*:\; \lambda\leq E_N({\bf k})\leq a_N\}.$$
We now choose $\delta$ small enough such that
$$\sigma(h({\bf k}_0))\cap [a_N-\delta,a_N+\delta]=\{E_N({\bf k}_0)\}.$$
We use analytic
perturbation theory in order to control the location of the spectrum
of $h({\bf k})$ when $\vert{\bf k}-{\bf k}_0\vert$ is small (we assume without
loss of generality that ${\bf k}_0$ lies in the interior of
$\Omega^*$). By writing
$$h({\bf k})=h({\bf k}_0) + ({\bf k}-{\bf k}_0)\cdot (-i\nabla +{\bf
  k}_0)+({\bf k}-{\bf k}_0)^2/2=:h({\bf k}_0)+W({\bf k}),$$
we see that we can find a constant $C>0$ such that
$$\Vert W({\bf k})(h({\bf k}_0)-i)^{-1}\Vert\leq C \vert{\bf k}-{\bf k}_0\vert,\quad
\vert{\bf k}-{\bf k}_0 \vert\leq 1.$$
Take a circle $\gamma$ with center at $a_N$ and radius
$r=(a_N-\lambda)/2\leq \delta/2$. For any $z\in\gamma$, by virtue of the first resolvent equation:
$$(h({\bf k}_0)-z)^{-1}=(h({\bf k}_0)-i)^{-1}+(z-i)(h({\bf
  k}_0)-i)^{-1}
(h({\bf k}_0)-z)^{-1}$$
and by using the estimate
$\Vert(h({\bf k}_0)-z)^{-1}\Vert =2/(a_N-\lambda)$, we can find
another constant $C_{\delta}>0$ such that:
$$\sup_{z\in\gamma}\Vert W({\bf k})(h({\bf k}_0)-z)^{-1}\Vert\leq C_{\delta}\frac{\vert{\bf k}-{\bf
    k}_0\vert}{(a_N-\lambda)},\quad
\vert{\bf k}-{\bf k}_0\vert\leq 1.$$
It turns out that if $\vert{\bf k}-{\bf
    k}_0\vert/(a_N-\lambda)$ is smaller than some $\epsilon>0$, then
$$ \sup_{z\in\gamma}\Vert W({\bf k})(h({\bf k}_0)-z)^{-1}\Vert\leq \epsilon C_{\delta}\; {\rm
  whenever}\;  \vert{\bf k}-{\bf k}_0\vert\leq (a_N-\lambda)\epsilon.$$
Standard analytic perturbation theory insures that
if $\epsilon$ is chosen small enough, $h({\bf k})$ will have
exactly $M$ eigenvalues inside $\gamma$. Thus for all $\mathbf{k}$ satisfying
$\vert \mathbf{k} - \mathbf{k}_{0}\vert \leq \epsilon (a_{N} -\lambda)$, we have
$\sigma(h(\mathbf{k})) \cap [a_{N} - \delta, a_{N}] \subseteq [(a_{N}+\lambda)/2, a_{N}] \subset [\lambda, a_{N}]$. In
particular, $\lambda <E_N({\bf k})\leq a_N$ for all such ${\bf
  k}$'s. But the ball in $\Omega^*$ where
$\vert{\bf k}-{\bf k}_0\vert\leq (a_N-\lambda)\epsilon$ has a volume which goes like
$(a_N-\lambda)^3$, and the proof is finished.

\qed

\vspace{0.5cm}
\subsection{Proof of Theorem \ref{metalsisemi}}

In this paragraph we prove the existence of the Fermi energy.
We separately investigate the semiconducting case and the metallic case.

\subsubsection{The semiconducting case (SC)}
We here consider the same situation as in Lemma \ref{ninr} in which there exists $N \geq 1$
such that $n_{\infty}(E) = \rho_{0}$ for all $E$ satisfying
$\max \mathcal{E}_{N} \leq E \leq \min \mathcal{E}_{N+1}$.
We set $a_{N} := \max \mathcal{E}_{N}$ and $b_{N} := \min
\mathcal{E}_{N+1}$.  Let $\mu(\beta):=\mu_{\infty}(\beta,\rho_{0})$ be
the unique solution of the equation $\rho_{\infty}(\beta, \mathrm{e}^{\beta \mu},0) = \rho_{0}$.
We start with the following lemma:
\begin{lema}\label{intreele}
\begin{equation}
\label{bornmu}
a_{N} \leq \mu_{1} := \liminf_{\beta \rightarrow  \infty} \mu(\beta) \leq \limsup_{\beta \rightarrow  \infty} \mu(\beta) =:\mu_{2} \leq b_{N}.
\end{equation}
\end{lema}
\noindent\textbf{\textit{Proof}}. We will only prove the inequality $a_N\leq \mu_1$, since the
proof of the other one ($\mu_2\leq b_N$) is similar. Assume the contrary: $\mu_1<a_N$. Define
$\epsilon:=a_N-\mu_1>0$. Then there exists a sequence
$\{\beta_n\}_{n\geq 1}$ with $\beta_n\rightarrow \infty$ and an integer $M_\epsilon \geq 1$
large enough such that:
\begin{equation*}
\lim_{n \rightarrow \infty} \mu(\beta_n) = \mu_1 \quad \textrm{and} \quad \mu(\beta_n)\leq
a_N-\epsilon/2<a_N, \quad \forall\, n \geq M_{\epsilon}.
\end{equation*}
Since $\rho_{\infty}(\beta,\mathrm{e}^{\beta \mu},0)$ is an increasing function of $\mu$,
we have:
\begin{equation*}
\rho_{0} = \rho_{\infty}(\beta_n,e^{\beta_n\mu(\beta_n)},0)
\leq \rho_{\infty}(\beta_n, e^{\beta_n(a_{N} - \epsilon/2)},0).
\end{equation*}
By letting $n \rightarrow \infty$ in the above inequality,
\eqref{limitrho} implies:
\begin{equation*}
\rho_{0} \leq n_{\infty}(a_{N} - \epsilon/2)<n_\infty(a_N)=\rho_0
\end{equation*}
where in the second inequality we used  \eqref{idsinf}.
We have arrived at a contradiction.
\qed

\vspace{0.5cm}

Now if $a_N=b_N$, the proof of \eqref{gapfermi} is over. Thus we can assume
that $a_N<b_N$, i.e. the gap is open. We have the following lemma:
\begin{lema}\label{punctfix0}
 Define $c_N=(a_N+b_N)/2$. For any $0<\epsilon<(b_N-a_N)/2$, there exists $\beta_\epsilon>0$ large enough such that
$\mu(\beta)\in [c_N-\epsilon,c_N+\epsilon]$ whenever $\beta>\beta_\epsilon$.
\end{lema}
\noindent\textbf{\textit{Proof}}. We know that $\mu(\beta)$ exists and is unique, thus if we can construct such a solution in the given interval, it means that this is the one. We use \eqref{rho3} in which we introduce $\mu(\beta)$ and
arrive at the following identities:
\begin{align*}
 n_{\infty}(a_{N}) &=\rho_{0}=\int_{- \infty}^{\infty} d\lambda \frac{\partial \mathfrak{f}_{FD}}
{\partial \lambda}
(\beta,\mu(\beta);\lambda)n_{\infty}(\lambda)
=- \int_{- \infty}^{a_{N}}  d\lambda\frac{\partial \mathfrak{f}_{FD}}
{\partial \lambda}
(\beta,\mu(\beta);\lambda)n_{\infty}(\lambda) \\
&- n_{\infty}(b_{N})\mathfrak{f}_{FD}(\beta,\mu(\beta);b_{N})
+
n_{\infty}(a_{N})\mathfrak{f}_{FD}(\beta,\mu(\beta);a_{N})
\\
&-
\int_{b_{N}}^{\infty}  d\lambda \frac{\partial \mathfrak{f}_{FD}}
{\partial \lambda} (\beta,\mu(\beta);\lambda) n_{\infty}(\lambda),
\end{align*}
where in the last term we used the fact that $n_\infty(\cdot)$ is constant on
the interval $[a_N,b_N]$, and this constant is nothing but $\rho_0$. We
can rewrite the above equation as:
\begin{equation}
\label{iddenti1}
\int_{- \infty}^{a_{N}} d\lambda \frac{\partial \mathfrak{f}_{FD}}{\partial
  \lambda}(\beta,\mu(\beta);\lambda)\{
n_{\infty}(\lambda)-n_{\infty}(a_{N}) \} = \int_{b_{N}}^{\infty} d\lambda
\frac{\partial \mathfrak{f}_{FD}}{\partial \lambda}
(\beta,\mu(\beta);\lambda)\{n_{\infty}(b_{N})-n_{\infty}(\lambda)\}
\end{equation}
where we used the fact that
$\mathfrak{f}_{FD}(\beta,\mu(\beta);-\infty)=1$ and
$\mathfrak{f}_{FD}(\beta,\mu(\beta);\lambda)\leq
Ce^{-\lambda \beta}$ for large $\lambda$.\\
In the left hand side of \eqref{iddenti1} we now introduce the explicit formula:
\begin{equation*}\label{dublaprostie4}
\partial_{\lambda}\mathfrak{f}_{FD}(\beta,\mu(\beta);\lambda) = -\beta \frac{\mathrm{e}^{\beta(\lambda -
    \mu(\beta))}}{(\mathrm{e}^{\beta(\lambda -
    \mu(\beta))}+ 1)^{2}} = - \beta
\mathrm{e}^{\beta(a_{N} - \mu(\beta))}
\frac{\mathrm{e}^{\beta(\lambda -
    a_{N})}}{(\mathrm{e}^{\beta(\lambda - \mu(\beta))}+ 1)^{2}},
\end{equation*}
while in the right hand side of \eqref{iddenti1} we use another expression:
\begin{equation*}\label{dublaprostie5}
\partial_{\lambda}\mathfrak{f}_{FD}(\beta,\mu(\beta);\lambda) = -\beta\frac{\mathrm{e}^{-\beta(\lambda-
    \mu(\beta))}}{( 1+\mathrm{e}^{-\beta(\lambda -
    \mu(\beta))})^{2}}
= - \beta
\mathrm{e}^{-\beta(b_{N} - \mu(\beta))}
\frac{\mathrm{e}^{-\beta(\lambda -
    b_{N})}}{( 1+\mathrm{e}^{-\beta(\lambda - \mu(\beta))})^{2}}.
\end{equation*}
Then \eqref{iddenti1} can be rewritten as:
\begin{align}
&\label{iddenti2}
\int_{- \infty}^{a_{N}} d\lambda
\frac{\mathrm{e}^{\beta(\lambda - a_{N})}}
{(\mathrm{e}^{\beta(\lambda - \mu(\beta))}+ 1)^{2}}
\{n_{\infty}(a_{N}) - n_{\infty}(\lambda)\} \nonumber \\
&= \mathrm{e}^{\beta\{2\mu(\beta) - (a_{N} + b_{N})\}}
\int_{b_{N}}^{\infty} d\lambda \frac{\mathrm{e}^{-\beta(\lambda -b_{N})}}
{(1 + \mathrm{e}^{-\beta(\lambda - \mu(\beta))})^{2}}
\{n_{\infty}(\lambda) - n_{\infty}(b_{N})\},
\end{align}
or by taking the logarithm:
\begin{align*}
&\mu(\beta)=c_N+\frac{1}{2\beta}\bigg \{\ln \bigg (\int_{- \infty}^{a_{N}} d\lambda
\frac{\mathrm{e}^{\beta(\lambda - a_{N})}}
{(\mathrm{e}^{\beta(\lambda - \mu(\beta))}+ 1)^{2}}
\{n_{\infty}(a_{N}) - n_{\infty}(\lambda)\} \bigg) \\
&-\ln \bigg(
\int_{b_{N}}^{\infty} d\lambda \frac{\mathrm{e}^{-\beta(\lambda -b_{N})}}
{(1 + \mathrm{e}^{-\beta(\lambda - \mu(\beta))})^{2}}
\{n_{\infty}(\lambda) - n_{\infty}(b_{N})\}\bigg )\bigg \}.
\end{align*}
Let us define the smooth function $f:[c_N-\epsilon,c_N+\epsilon]\mapsto \mathbb{R}$ given by:
\begin{align}\label{iddenti201}
f(x)&:=c_N+\frac{1}{2\beta}\bigg \{\ln \bigg (\int_{- \infty}^{a_{N}} d\lambda
\frac{\mathrm{e}^{\beta(\lambda - a_{N})}}
{(\mathrm{e}^{\beta(\lambda - x)}+ 1)^{2}}
\{n_{\infty}(a_{N}) - n_{\infty}(\lambda)\} \bigg)\nonumber \\
&-\ln \bigg(
\int_{b_{N}}^{\infty} d\lambda \frac{\mathrm{e}^{-\beta(\lambda -b_{N})}}
{(1 + \mathrm{e}^{-\beta(\lambda - x)})^{2}}
\{n_{\infty}(\lambda) - n_{\infty}(b_{N})\}\bigg)\bigg \}.
\end{align}
We will prove that if $\beta$ is large enough, then $f$ invariates the interval $[c_N-\epsilon,c_N+\epsilon]$, which is already enough for the existence of a fixed point. This would also show that $\mu(\beta)$ must be in that interval. But in fact one can prove more: $f$ is a contraction for large enough $\beta$.

The idea is to find some good upper and lower bounds when $\beta$ is large for the integrals under the logarithms.  We start by finding a lower bound in $\beta$ for the first integral. Let $\delta >0$ sufficiently
small. Using \eqref{idsinf} in the
left hand side of \eqref{iddenti2} we get:
\begin{equation}
\label{low1}\int_{- \infty}^{a_{N}} d\lambda
\frac{\mathrm{e}^{\beta(\lambda - a_{N})}}
{(\mathrm{e}^{\beta(\lambda - x)}+ 1)^{2}}
\{n_{\infty}(a_{N}) - n_{\infty}(\lambda)\}
\geq \frac{C}{4} \int_{a_{N}-\delta}^{a_{N}} \mathrm{e}^{- \beta(a_{N} -
  \lambda)}
(a_{N} - \lambda)^{3}
\end{equation}
where we used that $x\geq a_N\geq \lambda$ in order to
get rid of the numerator. After a change of variables and using some basic estimates
one arrives at another constant $C>0$ such that for $\beta$ sufficiently large:
\begin{equation}
\label{lowbound}
\int_{- \infty}^{a_{N}} d\lambda
\frac{\mathrm{e}^{\beta(\lambda - a_{N})}}
{(\mathrm{e}^{\beta(\lambda - x)}+ 1)^{2}}
\{n_{\infty}(a_{N}) - n_{\infty}(\lambda)\}  \geq
\frac{C}{\beta^{5}}.
\end{equation}
By restricting the interval of integration to $[b_{N},b_{N}+\delta]$ and by using \eqref{idsinf22},  we obtain by the same method a similar lower bound for the second integral under
the logarithm. Moreover, using the Weyl asymptotics which says that
$n_{\infty}(\lambda)\sim \lambda^{\frac{3}{2}}$ for large $\lambda$
(see e.g. \cite{KW}), one can also get a power-like upper bound in $\beta$ for our two
integrals.

We deduce from these estimates that there exists a constant $C_\epsilon >0$ such that:
$$\sup_{x\in [c_N-\epsilon,c_N+\epsilon]}\vert f(x)-c_N\vert\leq \frac{C_\epsilon \ln(\beta)}{\beta},\quad \beta >1.$$
Thus if $\beta$ is large enough, $f$ invariates the interval. Being continuous, it must have a fixed point. Moreover, the derivative $f'(x)$ decays exponentially with $\beta$
uniformly in $x\in [c_N-\epsilon,c_N+\epsilon]$. It implies that if $\beta$ is large enough,
then $\Vert f'\Vert_\infty<1$, that is $f$ is a contraction.
\qed

\vspace{0.5cm}
\subsubsection{The metallic case (M)}

Consider the situation in which there exists a unique solution
$E_M$ of the equation $n_{\infty}(E_M) = \rho_{0}$, and this solution
lies in the interior of a Bloch band. In other words, there
exists (a possibly not unique) integer $N \geq 1$ such that
$\min \mathcal{E}_{N} < E_M < \max \mathcal{E}_{N}$. We will use in the following that the IDS
$n_\infty(\cdot)$ is a strictly increasing function on the interval
$[\min \mathcal{E}_{N}, \max \mathcal{E}_{N}]$.

Let $\mu(\beta):=\mu_{\infty}(\beta,\rho_{0})$ be the unique real solution
of the equation
$\rho_{\infty}(\beta,e^{\beta \mu(\beta)},0) = \rho_{0}$.
Let us show that :
\begin{equation}
E_M\leq \liminf_{\beta \rightarrow \infty} \mu(\beta) \leq
\limsup_{\beta \rightarrow  \infty} \mu(\beta)\leq E_M,
\end{equation}
which would end the proof. We start with the first inequality.

Assume ad-absurdum that $\mu_1:=\liminf_{\beta \rightarrow \infty}
\mu(\beta)<E_M$. Then there exists $\epsilon>0$ and a sequence
$\{\beta_n\}_{n \geq 1}$ satisfying $\beta_n \rightarrow \infty$ such that:
$$\lim_{n\to\infty}\mu(\beta_n)= \mu_1,\quad \mu(\beta_n)\leq
E_M-\epsilon,\quad \forall n\geq 1.$$
 Since $\rho_{\infty}(\beta,e^{\beta\mu},0)$ is increasing with $\mu$, we have:
\begin{equation}
n_\infty(E_M)=\rho_0=\lim_{n\to\infty}\rho_{\infty}(\beta_n,
e^{\beta_n\mu(\beta_n)},0)\leq \lim_{n\to\infty}\rho_{\infty}(\beta_n,
e^{\beta_n(E_M-\epsilon)},0)=n_\infty(E_M-\epsilon),
\end{equation}
where in the last equality we used \eqref{limitrho}.
But the inequality $n_\infty(E_M)\leq n_\infty(E_M-\epsilon)$
is in contradiction with the fact that $n_{\infty}(\cdot)$ is a
strictly increasing function near $E_M$. Thus $E_M\leq \mu_1$.

Now assume ad-absurdum that $\mu_2:=\limsup_{\beta \rightarrow \infty}
\mu(\beta)>E_M$. Then there exists $\epsilon>0$ and a sequence
$\{\beta_n\}_{n \geq 1}$ satisfying $\beta_n \rightarrow  \infty$ such that:
$$\lim_{n\to\infty}\mu(\beta_n)= \mu_2,\quad
E_M+\epsilon\leq \mu(\beta_n) ,\quad \forall n\geq 1.$$
We again use that $\rho_{\infty}(\beta,e^{\beta\mu},0)$ is increasing with $\mu$ and write:
\begin{equation}
n_\infty(E_M+\epsilon)=\lim_{n\to\infty}\rho_{\infty}(\beta_n,
e^{\beta_n(E_M+\epsilon)},0)\leq \lim_{n\to\infty}\rho_{\infty}(\beta_n,
e^{\beta_n\mu(\beta_n)},0)=\rho_0=n_\infty(E_M),
\end{equation}
where in the first equality we again used \eqref{limitrho}.
But the inequality $n_\infty(E_M+\epsilon)\leq n_\infty(E_M)$
is also in contradiction with the fact that $n_{\infty}(\cdot)$ is a
strictly increasing function near $E_M$. Therefore $\mu_2\leq E_M$.
\qed

\section{The zero-field susceptibility at fixed density and positive temperature}

In this section we prove a general formula for the zero-field grand-canonical
susceptibility of a Bloch electrons gas at fixed density and positive temperature.

Here is the main result of this section:

\begin{theorem}
\label{genexpand}
Let $\beta >0$ and $\rho_{0}>0$ be fixed. Let $\mu_{\infty}=\mu_{\infty}(\beta,\rho_{0}) \in \mathbb{R}$ the unique solution of the equation
$\rho_{\infty}(\beta,\mathrm{e}^{\beta \mu},\omega=0) = \rho_{0}$.
Then for each integer $j_{1} \geq 1$ there exists four families of functions
$\mathfrak{c}_{j_{1},l}(\cdot\,)$, with $l \in \{0,1,2,3\}$, defined on $\Omega^{*}$ outside
a set of Lebesgue measure zero, such that the integrand below can be extended by
continuity to the whole of $\Omega^{*}$:
\begin{equation}
\label{exformula2}
\mathcal{X}(\beta,\rho_{0}) = - \bigg(\frac{e}{c}\bigg)^{2}
\frac{1}{2\beta}
\frac{1}{(2\pi)^3} \sum_{j_{1}=1}^{\infty} \int_{\Omega^{*}}
\mathrm{d}\mathbf{k}\, \sum_{l= 0}^{3} \frac{\partial^{l} \mathfrak{f}}{\partial \xi^{l}}\big(\beta,\mu_{\infty};E_{j_{1}}(\mathbf{k})\big) \mathfrak{c}_{j_{1},l}(\mathbf{k}),
\end{equation}
with the convention $(\partial_{\xi}^{0} \mathfrak{f})(\beta, \mu_{\infty}; E_{j_{1}}(\mathbf{k})) = \mathfrak{f}(\beta,\mu_{\infty};E_{j_{1}}(\mathbf{k})) := \ln(1 + \mathrm{e}^{\beta(\mu_{\infty} - E_{j_{1}}(\mathbf{k}))})$.
\end{theorem}
\vspace{0.5cm}

This formula is a necessary step in the proof of Theorem \ref{maintheorem} $(i)$ and $(ii)$ (this is the aim of the following section) when we will take the limit of zero temperature.

The special feature of this formula lies in the fact that each function
$\mathfrak{c}_{j_{1},l}(\cdot)$ can be only expressed in terms of Bloch energy functions
and their associated eigenfunctions. For each integer $j_{1} \geq 1$, the functions $\mathfrak{c}_{j_{1},2}(\cdot)$ and $\mathfrak{c}_{j_{1},3}(\cdot)$ are identified respectively in
\eqref{c2} and \eqref{c3}. As for the functions $\mathfrak{c}_{j_{1},l}(\cdot)$ with
$l \in \{0,1\}$, they can also be written down but their explicit expression is not
important  for the proof of Theorem \ref{maintheorem}. Note as well that the above formula
brings into play the Fermi-Dirac distribution and its partial derivatives up to
the second order. This will turn out to be very important when we will take the limit
$\beta \rightarrow \infty$ in the following section.
\subsection{Starting the proof: a general formula from the magnetic perturbation theory}

We start by giving a useful formula for the thermodynamic limit of the
grand-canonical susceptibility. Let $\beta > 0$ and $z :=
\mathrm{e}^{\beta \mu} \in (0,\infty)$ the fixed external
parameters. Let $\Gamma$ be the positively oriented contour defined in
\eqref{contour}, going round the half-line $[E_{0},\infty)$, and included in the
analiticity domain of the map
$\xi \mapsto \mathfrak{f}(\beta,z;\xi) = \ln(1 + z \mathrm{e}^{- \beta \xi})$.
Denote by $R_{\infty}(\omega,\xi) :=
(H_{\infty}(\omega) - \xi)^{-1}$ for all $\xi \in
\rho(H_{\infty}(\omega))$ and $\omega \in \mathbb{R}$.
Taking into account the periodic structure of our
  system, it is proved (see \cite{BCS1}, Theorem 3.8) that the
  thermodynamic limit of the grand-canonical pressure of the Bloch electron gas
at any intensity of the magnetic field $B$ is given by:
\begin{equation}
\label{Pinfini}
P_{\infty}(\beta,z,\omega) := \frac{1}{\beta \vert \Omega \vert}
\frac{i}{2\pi}
\mathrm{Tr}_{L^{2}(\mathbb{R}^{3})}\bigg\{\chi_{\Omega} \int_{\Gamma}
\mathrm{d}\xi\,
\mathfrak{f}(\beta,z;\xi) R_{\infty}(\omega,\xi)\bigg\},
\end{equation}
where $\Omega$ is the unit cube centered at the origin of coordinates
($\chi_{\Omega}$ denotes its characteristic function).
Although the integral kernel $R_{\infty}(\cdot\,,\cdot\,;\omega,\xi)$
of the resolvent
 has a singularity on the diagonal, the integration with respect to
$\xi$ in \eqref{Pinfini} provides us with a jointly continuous kernel
on $\mathbb{R}^{3}\times\mathbb{R}^{3}$. One can see this by performing an
integration by parts in \eqref{Pinfini} and using the fact that the
kernel of $R_{\infty}^{2}(\omega,\xi)$ is jointly continuous.
Moreover, one can prove \cite{BrCoLo1, BrCoLo2, BrCoLo3}
that the thermodynamic limit of the grand-canonical pressure
is jointly smooth on
$(z,\omega)\in (- \mathrm{e}^{\beta E_{0}},\infty)\times
\mathbb{R}$.

Let $\omega \in \mathbb{R}$ and $\xi \in
\rho(H_{\infty}(\omega))$.
Introduce the bounded operators $T_{\infty,1}(\omega,\xi)$ and
$T_{\infty,2}(\omega,\xi)$ generated by the following integral kernels:
\begin{align}
\label{Sinfini1}
T_{\infty,1}(\mathbf{x},\mathbf{y};\omega,\xi) &:=
\mathbf{a}(\mathbf{x} - \mathbf{y}) \cdot (i\nabla_{\mathbf{x}} + \omega\mathbf{a}(\mathbf{x})) R_{\infty}(\mathbf{x},\mathbf{y};\omega,\xi) \\
\label{Sinfini2}
T_{\infty,2}(\mathbf{x},\mathbf{y};\omega,\xi) &:= \frac{1}{2}
\mathbf{a}^{2}(\mathbf{x} - \mathbf{y})
R_{\infty}(\mathbf{x},\mathbf{y};\omega,\xi), \quad \mathbf{x}\neq \mathbf{y},
\end{align}
where $\mathbf{a}(\cdot)$ stands for the usual symmetric gauge
$\mathbf{a}(\mathbf{x}) = \frac{1}{2} \mathbf{e}_{3} \wedge \mathbf{x} =
\frac{1}{2}(-x_{2},x_{1},0)$.
We introduce the following operators :
\begin{align}
\label{Winfini1}
\mathcal{W}_{\infty,1}(\beta,\mu,\omega) &:=
\frac{i}{2\pi}\int_{\Gamma}
\mathrm{d}\xi\,\mathfrak{f}(\beta,\mu;\xi)R_{\infty}(\omega,\xi)
T_{\infty,1}(\omega,\xi) T_{\infty,1}(\omega,\xi) \\
\label{Winfini2}
\mathcal{W}_{\infty,2}(\beta,\mu,\omega) &:= \frac{i}{2\pi}
\int_{\Gamma}
\mathrm{d}\xi\,\mathfrak{f}(\beta,\mu;\xi)R_{\infty}(\omega,\xi)T_{\infty,2}
(\omega,\xi)
\end{align}
One can prove using the same techniques as in \cite{CN3} that these
operators are locally trace class and have a jointly continuous kernel on
$\mathbb{R}^{3}\times\mathbb{R}^{3}$. By a closely
  related method as the one in \cite{BrCoLo1}, \cite{BrCoLo2}, it is
  proved in \cite{BCS2} that we can invert the thermodynamic limit
with the partial derivatives w.r.t. $\omega$ of the grand-canonical
pressure. Then the bulk orbital susceptibility reads as:
\begin{equation*}
\label{chiGC}
\begin{split}
\mathcal{X}_{\infty}^{GC}(\beta,e^{\beta\mu},\omega) &:=
\bigg(\frac{e}{c}\bigg)^{2} \frac{\partial^{2} P_{\infty}}{\partial
  \omega^{2}}
(\beta,e^{\beta\mu},\omega) \\
&= \bigg(\frac{e}{c}\bigg)^{2} \frac{2}{\beta \vert \Omega \vert}
\Big\{\mathrm{Tr}_{L^{2}(\mathbb{R}^{3})} \big\{\chi_{\Omega}
\mathcal{W}_{\infty,1}(\beta,\mu,\omega)\big\} -
\mathrm{Tr}_{L^{2}(\mathbb{R}^{3})} \big\{\chi_{\Omega}
\mathcal{W}_{\infty,2}(\beta,\mu,\omega)\big\}\Big\}
\end{split}
\end{equation*}
We mention that the above formula is obtained using the
  so-called 'gauge invariant magnetic perturbation theory' applied to
  the resolvent integral kernel (see e.g. \cite{CN3} for further
  details) which allows to control the linear growth of the magnetic vector potential.

The quantity which we are interested in is the orbital
susceptibility
at zero magnetic field and at fixed density of particles $\rho_{0}$.
Note that the pressure is an even function of
$\omega$, thus its first order derivative at $\omega=0$ is zero. This
explains why the susceptibility is the relevant physical quantity for
the weak magnetic field regime.

The orbital susceptibility at zero magnetic field and fixed
density $\rho_{0}$ is given by (see also \eqref{muinfini}):
\begin{equation}
\label{chiC}
\begin{split}
\mathcal{X}(\beta,\rho_{0}) :&=
\mathcal{X}_{\infty}^{GC}(\beta,e^{\beta\mu_{\infty}(\beta,\rho_{0})},0) \\
&= \bigg(\frac{e}{c}\bigg)^{2} \frac{2}{\beta \vert \Omega \vert}
\Big\{\mathrm{Tr}_{L^{2}(\mathbb{R}^{3})} \big\{\chi_{\Omega}
\mathcal{W}_{\infty,1}(\beta,\mu_{\infty},0)\big\} -
\mathrm{Tr}_{L^{2}(\mathbb{R}^{3})} \big\{\chi_{\Omega}
\mathcal{W}_{\infty,2}(\beta,\mu_{\infty},0)\big\}\Big\}.
\end{split}
\end{equation}

The formula \eqref{chiC} constitutes the starting-point in obtaining
\eqref{exformula2}. The next step consists in rewriting the local traces appearing
in \eqref{chiC} in a more convenient way:
\begin{proposition}
\label{propo1}
Let $p_{\alpha}:=-i\partial_{\alpha}$ with $\alpha \in \{1,2,3\}$ be
the cartesian
components of the momentum operator defined in $L^2(\mathbb{R}^3)$.
Then we have:
\begin{multline}
\label{term1}
\mathrm{Tr}_{L^{2}(\mathbb{R}^{3})}\big\{\chi_{\Omega}\mathcal{W}_{\infty,1}(\beta,\mu_{\infty},0)\big\}
= \frac{1}{4}\frac{i}{2\pi}
\mathrm{Tr}_{L^{2}(\mathbb{R}^{3})}\bigg\{\chi_{\Omega}\int_{\Gamma}
\mathrm{d}\xi\, \mathfrak{f}(\beta,\mu_{\infty};\xi) \\
\big[R_{\infty}(0,\xi)p_{1}R_{\infty}(0,\xi)p_{2}R_{\infty}(0,\xi)\big\{p_{2}R_{\infty}(0,\xi)p_{1}
R_{\infty}(0,\xi) - p_{1}R_{\infty}(0,\xi)p_{2}R_{\infty}(0,\xi)\big\} + \\
+
R_{\infty}(0,\xi)p_{2}R_{\infty}(0,\xi)p_{1}R_{\infty}(0,\xi)\big\{p_{1}R_{\infty}(0,\xi)p_{2}R_{\infty}(0,\xi)
- p_{2}R_{\infty}(0,\xi)p_{1}R_{\infty}(0,\xi)\big\}\big]
\bigg\}
\end{multline}
and
\begin{multline}
\label{term2}
\mathrm{Tr}_{L^{2}(\mathbb{R}^{3})}\big\{\chi_{\Omega}\mathcal{W}_{\infty,2}(\beta,\mu_{\infty},0)\big\} = -\frac{1}{4}\frac{i}{2\pi} \mathrm{Tr}_{L^{2}(\mathbb{R}^{3})}\bigg\{\chi_{\Omega} \int_{\Gamma} \mathrm{d}\xi\, \mathfrak{f}(\beta,\mu_{\infty};\xi) \\
 R_{\infty}(0,\xi)R_{\infty}(0,\xi)\big[p_{2}R_{\infty}(0,\xi)p_{2}R_{\infty}(0,\xi) + p_{1}
R_{\infty}(0,\xi)p_{1}R_{\infty}(0,\xi) - R_{\infty}(0,\xi)\big]\bigg\}.
\end{multline}
\end{proposition}
\noindent \textbf{\textit{Proof}}. We begin with the justification of \eqref{term2}.
By rewriting \eqref{Sinfini2} as:
\begin{equation*}
\begin{split}
T_{\infty,2}(\mathbf{x},\mathbf{y};0,\xi) &= \frac{1}{8}
\{\mathbf{e}_{3}\wedge(\mathbf{x} - \mathbf{y})\}\cdot\{
\mathbf{e}_{3}\wedge(\mathbf{x} -\mathbf{y})\}R_{\infty}(\mathbf{x},\mathbf{y};0,\xi)\\
 &= \frac{1}{8}\big[(x_{2} - y_{2})^{2}+(x_{1} - y_{1})^{2}\big] R_{\infty}(\mathbf{x},\mathbf{y};0,\xi),
\end{split}
\end{equation*}
from \eqref{Winfini2} it follows:
\begin{align}
\label{eq1}
&\mathcal{W}_{\infty,2}(\mathbf{x},\mathbf{x};\beta,\mu,0)\\
&= \frac{1}{8}\int_{\Gamma} \mathrm{d}\xi\,\mathfrak{f}(\beta,\mu;\xi)
\int_{\mathbb{R}^{3}}
\mathrm{d}\mathbf{z}\,R_{\infty}(\mathbf{x},\mathbf{z};0,\xi) \big[(z_{2} -
x_{2})^{2}+(z_{1} - x_{1})^{2}\big]
R_{\infty}(\mathbf{z},\mathbf{x};0,\xi), \quad \forall\,\mathbf{x} \in \mathbb{R}^{3}.\nonumber
\end{align}
Let $l \in \{1,2\}$. Denote by $\mathbf{X}$ the multiplication operator
with $\mathbf{x}$. Then for all $\mathbf{z}\neq \mathbf{x}$ we can write:
\begin{equation*}\label{prostye2}
(z_{l} - x_{l}) R_{\infty}(\mathbf{z},\mathbf{x};0,\xi) = \big[\mathbf{X}\cdot\mathbf{e}_{l},R_{\infty}(0,\xi)\big](\mathbf{z},\mathbf{x}) = \big\{R_{\infty}(0,\xi)\big[H_{\infty}(0),\mathbf{X}\cdot\mathbf{e}_{l}\big]R_{\infty}(0,\xi)\big\}(\mathbf{z},\mathbf{x}).
\end{equation*}
We know that $[H_{\infty}(0),\mathbf{X}\cdot\mathbf{e}_{l}] = - i p_{l}$. Thus:
\begin{equation}
\label{eq2}
(z_{l} - x_{l})R_{\infty}(\mathbf{z},\mathbf{x};0,\xi) = -i
\big\{R_{\infty}(0,\xi)p_{l}R_{\infty}(0,\xi)\big\}(\mathbf{z},\mathbf{x}).
\end{equation}
Using standard commutation rules, we deduce from \eqref{eq2} that for
$l \in \{1,2\}$ and for all $\mathbf{z}\neq\mathbf{x}$:
\begin{equation}
\label{eq3}
(z_{l} - x_{l})^{2} R_{\infty}(\mathbf{z},\mathbf{x};0,\xi) = -\big\{2 R_{\infty}(0,\xi)p_{l}R_{\infty}(0,\xi)p_{l}R_{\infty}(0,\xi) - R_{\infty}(0,\xi)R_{\infty}(0,\xi)\big\}(\mathbf{z},\mathbf{x}).
\end{equation}
It remains to put \eqref{eq3} in \eqref{eq1}, and we get
\eqref{term2}.

Let us now prove now \eqref{term1}. Since the divergence of $\mathbf{a}$
is zero, then for $\mathbf{x}\neq \mathbf{y}$ we have:
\begin{align*}
T_{\infty,1}(\mathbf{x},\mathbf{y};0,\xi) &=  \frac{i}{2} \nabla_{\mathbf{x}} \cdot \{
\mathbf{e}_{3} \wedge (\mathbf{x} - \mathbf{y}) \}R_{\infty}(\mathbf{x},\mathbf{y};0,\xi)\\
&= i \nabla_{\mathbf{x}} \cdot \Big[- \frac{(x_{2} - y_{2})}{2} \mathbf{e}_{1} + \frac{(x_{1} - y_{1})}{2} \mathbf{e}_{2}\Big] R_{\infty}(\mathbf{x},\mathbf{y};0,\xi).
\end{align*}
From \eqref{Winfini1} it follows that for all $\mathbf{x} \in \mathbb{R}^{3}$:
\begin{multline*}
\mathcal{W}_{\infty,1}(\mathbf{x},\mathbf{x};\beta,\mu,0) = \frac{1}{4} \int_{\Gamma} \mathrm{d}\xi\, \mathfrak{f}(\beta,\mu;\xi) \int_{\mathbb{R}^{3}} \mathrm{d}\mathbf{z}_{1} \int_{\mathbb{R}^{3}} \mathrm{d}\mathbf{z}_{2}\,R_{\infty}(\mathbf{x},\mathbf{z}_{1};0,\xi) \\  \big\{(i \nabla_{\mathbf{z}_{1}} \cdot \mathbf{e}_{1})[-(z_{1,2} - z_{2,2}) R_{\infty}(0,\xi)(\mathbf{z}_{1},\mathbf{z}_{2})] + (i \nabla_{\mathbf{z}_{1}} \cdot \mathbf{e}_{2})[(z_{1,1} - z_{2,1}) R_{\infty}(0,\xi)(\mathbf{z}_{1},\mathbf{z}_{2})]\big\} \cdot\\
\cdot \big\{(i \nabla_{\mathbf{z}_{2}} \cdot \mathbf{e}_{1})[-(z_{2,2} - x_{2}) R_{\infty}(0,\xi)(\mathbf{z}_{2},\mathbf{x})] + (i \nabla_{\mathbf{z}_{2}} \cdot \mathbf{e}_{2})[(z_{2,1} - x_{1}) R_{\infty}(0,\xi)(\mathbf{z}_{2},\mathbf{x})]\big\}.
\end{multline*}
Then by using \eqref{eq2}, we get \eqref{term1} from the following identity:
\begin{multline*}
\forall\,\mathbf{x} \in \mathbb{R}^{3},\quad \mathcal{W}_{\infty,1}(\mathbf{x},\mathbf{x};\beta,\mu,0) = \frac{1}{4} \int_{\Gamma} \mathrm{d}\xi\, \mathfrak{f}(\beta,\mu;\xi) \int_{\mathbb{R}^{3}} \mathrm{d}\mathbf{z}_{1} \int_{\mathbb{R}^{3}} \mathrm{d}\mathbf{z}_{2}\,R_{\infty}(\mathbf{x},\mathbf{z}_{1};0,\xi) \\ \big\{ip_{1}\big(R_{\infty}(0,\xi)p_{2}R_{\infty}(0,\xi)\big)(\mathbf{z}_{1},\mathbf{z}_{2}) - ip_{2}\big(R_{\infty}(0,\xi)p_{1}R_{\infty}(0,\xi)\big)(\mathbf{z}_{1},\mathbf{z}_{2})\big\} \\ \big\{ip_{1}\big(R_{\infty}(0,\xi)p_{2}R_{\infty}(0,\xi)\big)(\mathbf{z}_{2},\mathbf{x}) - ip_{2}\big(R_{\infty}(0,\xi)p_{1}R_{\infty}(0,\xi)\big)(\mathbf{z}_{2},\mathbf{x})\big\}.
\end{multline*}
\qed

\subsection{Using the Bloch decomposition}

We know that (see e.g. \cite{BS}) $H_{\infty}(0)$ can be seen as a direct integral
$\int_{\Omega^{*}}^{\oplus} \mathrm{d}\mathbf{k}\, h(\mathbf{k})$ where the fiber Hamiltonians $h(\mathbf{k})$ acting in $L^{2}(\mathbb{T}^{3})$ are given by :
\begin{equation}
h(\mathbf{k}) = \frac{1}{2} (-i \nabla + \mathbf{k})^{2} + V.
\end{equation}
Recall that $h(\mathbf{k})$ is essentially self-adjoint in
$\mathcal{C}^{\infty}(\mathbb{T}^{3})$;
the domain of its closure is the Sobolev space $\mathcal{H}^{2}(\mathbb{T}^{3})$.
For each $\mathbf{k} \in \Omega^{*}$, $h(\mathbf{k})$ has purely discrete
spectrum. We have already denoted by $\{E_{j}(\mathbf{k})\}_{j \geq 1}$ the set of
eigenvalues counting
multiplicities and in increasing order. The corresponding
eigenfunctions
$\{u_{j}(\cdot\,;\mathbf{k})\}_{j \geq 1}$ form a
complete orthonormal system in $L^{2}(\mathbb{T}^{3})$ and satisfy:
\begin{equation*}
h(\mathbf{k}) u_{j}(\cdot;\mathbf{k}) = E_{j}(\mathbf{k}) u_{j}(\cdot;\mathbf{k}).
\end{equation*}
The eigenfunctions $u_j$'s are defined up to an
  arbitrary phase depending on $\mathbf{k}$. These phases cannot be
  always chosen to be continuous at crossing points, and even less
  differentiable.
For the following let us introduce another notation. Let $\alpha \in \{1,2,3\}$, and let
$i,j \geq 1$ be any natural numbers. Then for all $\mathbf{k} \in \Omega^{*}$ we define:
\begin{equation}
\label{hatpi}
\hat{\pi}_{i,j}(\alpha;\mathbf{k}) := \int_{\Omega}
\mathrm{d}{\mathbf{x}}\,
\overline{u_{i}({\mathbf{x}};\mathbf{k})}[(p_{\alpha} +
k_{\alpha})u_{j}({\mathbf{x}};\mathbf{k})]
= \langle u_{i}(\cdot\,;\mathbf{k}),(p_{\alpha} + k_{\alpha})u_{j}(\cdot\,;\mathbf{k})\rangle.
\end{equation}
Note that due to the phases presence in
the eigenfunctions $u_j$'s, we cannot be sure that the $\hat \pi_{i,j}$'s are
continuous/differentiable at crossing points. But all these 'bad'
phase factors will disappear when we take the traces
(see \eqref{retrace1'} and \eqref{retrace2'} below).

We now can write the local traces of Proposition \ref{propo1} in the following way:
\begin{proposition}
\label{propo2}
Let $\beta >0$ and $\rho_{0} >0$ be fixed. Let $\mu_{\infty} =
\mu_{\infty}(\beta,\rho_{0}) \in \mathbb{R}$ be the unique solution of the
equation $\rho_{\infty}(\beta,e^{\beta\mu},0) = \rho_{0}$.
Then both quantities \eqref{term1} and \eqref{term2} can be rewritten as:
\begin{multline}
\label{trace1'}
\mathrm{Tr}_{L^{2}(\mathbb{R}^{3})}\big\{\chi_{\Omega}
\mathcal{W}_{\infty,1}(\beta,\mu_{\infty},0)\big\}= -
\frac{1}{4}\frac{1}{\vert \Omega^{*} \vert}
\sum_{j_{1},\ldots,j_{4}=1}^{\infty}
\int_{\Omega^{*}} \mathrm{d}\mathbf{k}\, \mathcal{C}_{j_{1},j_{2},j_{3},j_{4}}(\mathbf{k}) \\
 \frac{1}{2i\pi}\int_{\Gamma} \mathrm{d}\xi\,
\frac{\mathfrak{f}(\beta,\mu_{\infty};\xi)}{\big(E_{j_{1}}(\mathbf{k}) -
  \xi\big)^{2}
\big(E_{j_{2}}(\mathbf{k}) - \xi\big)\big(E_{j_{3}}(\mathbf{k}) - \xi\big)
\big(E_{j_{4}}(\mathbf{k}) - \xi\big)},
\end{multline}
and
\begin{multline}
\label{trace2'}
\mathrm{Tr}_{L^{2}(\mathbb{R}^{3})}\big\{\chi_{\Omega} \mathcal{W}_{\infty,2}(\beta,\mu_{\infty},0)\big\} = - \frac{1}{4}\frac{1}{\vert \Omega^{*} \vert} \bigg\{ \sum_{j_{1}= 1}^{\infty} \int_{\Omega^{*}} \mathrm{d}\mathbf{k}\, \frac{1}{2i\pi}\int_{\Gamma} \mathrm{d}\xi\, \frac{\mathfrak{f}(\beta,\mu_{\infty};\xi)}{\big(E_{j_{1}}(\mathbf{k}) - \xi\big)^{3}} + \\
- \sum_{j_{1},j_{2}=1}^{\infty}  \int_{\Omega^{*}}
\mathrm{d}\mathbf{k}\,
\mathcal{C}_{j_{1},j_{2}}(\mathbf{k})
\frac{1}{2i\pi}\int_{\Gamma} \mathrm{d}\xi\,
\frac{\mathfrak{f}(\beta,\mu_{\infty};\xi)}
{\big(E_{j_{1}}(\mathbf{k}) - \xi\big)^{3}\big(E_{j_{2}}(\mathbf{k}) - \xi\big)}\bigg\},
\end{multline}
where the functions $\Omega^{*} \owns \mathbf{k} \mapsto \mathcal{C}_{j_{1},j_{2},j_{3},j_{4}}(\mathbf{k})$ and $\Omega^{*} \owns \mathbf{k} \mapsto \mathcal{C}_{j_{1},j_{2}}(\mathbf{k})$ are defined by:
\begin{multline}
\label{cj4}
\mathcal{C}_{j_{1},j_{2},j_{3},j_{4}}(\mathbf{k}) := \big\{ \hat{\pi}_{j_{1},j_{2}}(1;\mathbf{k}) \hat{\pi}_{j_{2},j_{3}}(2;\mathbf{k}) - \hat{\pi}_{j_{1},j_{2}}(2;\mathbf{k})\hat{\pi}_{j_{2},j_{3}}(1;\mathbf{k})\big\} \\\
\times \big\{ \hat{\pi}_{j_{3},j_{4}}(2;\mathbf{k}) \hat{\pi}_{j_{4},j_{1}}(1;\mathbf{k}) - \hat{\pi}_{j_{3},j_{4}}(1;\mathbf{k})\hat{\pi}_{j_{4},j_{1}}(2;\mathbf{k})\big\}
\end{multline}
and
\begin{equation}
\label{cj2}
\mathcal{C}_{j_{1},j_{2}}(\mathbf{k}) := \hat{\pi}_{j_{1},j_{2}}(1;\mathbf{k}) \hat{\pi}_{j_{2},j_{1}}(1;\mathbf{k}) + \hat{\pi}_{j_{1},j_{2}}(2;\mathbf{k})\hat{\pi}_{j_{2},j_{1}}(2;\mathbf{k}) = \big\vert \hat{\pi}_{j_{1},j_{2}}(1;\mathbf{k})\big\vert^{2} +  \big\vert \hat{\pi}_{j_{1},j_{2}}(2;\mathbf{k})\big\vert^{2}.
\end{equation}
\end{proposition}
\vspace{0.5cm}

We do not give more details since this result is just a straightforward application of the following rather non-trivial
technical lemma (recently proved in \cite{CN}):
\begin{lema}
\label{inversesum}
Let $\beta >0$ and $\mu \in \mathbb{R}$ be fixed. For $n,m \in \mathbb{N}$ with
$m,n\geq 1$, consider the local trace given by:
\begin{equation*}
\mathcal{J}_{\alpha_{1},\ldots,\alpha_{n}}^{(m)}:= \mathrm{Tr}_{L^{2}(\mathbb{R}^{3})} \bigg\{ \chi_{\Omega} \int_{\Gamma} \mathrm{d}\xi\, \mathfrak{f}(\beta,\mu;\xi) (H_{\infty}(0) - \xi)^{-m} p_{\alpha_{1}} (H_{\infty}(0) - \xi)^{-1}\dotsb p_{\alpha_{n}} (H_{\infty}(0) - \xi)^{-1}\bigg\}
\end{equation*}
Then under the assumption that $V \in
\mathcal{C}^{\infty}(\mathbb{T}^{3})$ we have:
\begin{multline}
\label{systemi}
\mathcal{J}_{\alpha_{1},\ldots,\alpha_{n}}^{(m)} =
\frac{1}{\vert \Omega^{*} \vert} \sum_{j_{1},\ldots,j_{n}\geq 1}
\int_{\Omega^{*}} \mathrm{d}\mathbf{k}\,
\hat{\pi}_{j_{1},j_{2}}(\alpha_{1};\mathbf{k})\dots
\hat{\pi}_{j_{n},j_{1}}(\alpha_{n};\mathbf{k}) \\
 \int_{\Gamma} \mathrm{d}\xi\, \frac{\mathfrak{f}(\beta,\mu;\xi)}{\big(E_{j_{1}}(\mathbf{k}) - \xi\big)^{m+1} \big(E_{j_{2}}(\mathbf{k}) - \xi\big) \dotsb \big(E_{j_{n}}(\mathbf{k}) - \xi\big)}.
\end{multline}
where all the above series are absolutely convergent and
$\hat{\pi}_{i,j}(\alpha;\mathbf{k})$ is defined by \eqref{hatpi}.
\end{lema}


\subsection{Applying the residue calculus}

Consider the expression of the susceptibility at fixed density \eqref{chiC} in which the
local traces are now given by
\eqref{trace1'} and \eqref{trace2'}. Remark that these quantities now are written in
a convenient way in order to apply the residue theorem. Denote the
integrands appearing in \eqref{trace1'} and \eqref{trace2'} by:
\begin{align*}
\mathfrak{g}_{j_{1},j_{2}}(\beta,\mu_{\infty};\xi) &:= \frac{\mathfrak{f}(\beta,\mu_{\infty};\xi)}{\big(E_{j_{1}}(\mathbf{k}) - \xi\big)^{3} \big(E_{j_{2}}(\mathbf{k}) - \xi\big)},\quad j_{1},j_{2} \in \mathbb{N}^{*} \\
\mathfrak{h}_{j_{1},j_{2},j_{3},j_{4}}(\beta,\mu_{\infty};\xi) &:= \frac{\mathfrak{f}(\beta,\mu_{\infty};\xi)}{\big(E_{j_{1}}(\mathbf{k}) - \xi\big)^{2} \big(E_{j_{2}}(\mathbf{k}) - \xi\big)\big(E_{j_{3}}(\mathbf{k}) - \xi\big)\big(E_{j_{4}}(\mathbf{k}) - \xi\big)},\quad j_{1},j_{2},j_{3},j_{4}\in \mathbb{N}^{*}.
\end{align*}
Note that $\mathfrak{g}_{j_{1},j_{2}}(\beta,\mu_{\infty};\cdot\,)$ can
have first order, third order, or even fourth order poles (in the case
when $j_{1} = j_{2}$). In the same way,
$\mathfrak{h}_{j_{1},j_{2},j_{3},j_{4}}(\beta,\mu_{\infty};\cdot\,)$
can have poles from the first order up to at most fifth order (in the
case when $j_{1} = j_{2} = j_{3} = j_{4}$). Hence we expect that the integrals of
$\mathfrak{h}_{j_{1},j_{2},j_{3},j_{4}}(\beta,\mu_{\infty};\cdot\,)$
in \eqref{trace1'}
(resp. of $\mathfrak{g}_{j_{1},j_{2}}(\beta,\mu_{\infty};\cdot\,)$ in
\eqref{trace2'}) to make appear partial derivatives of $\mathfrak{f}(\beta,\mu_{\infty};\cdot\,)$ with
order at most $4$ (resp. with order at most $3$). But we will see
below that the factor multiplying
$(\partial_{\xi}^{4}\mathfrak{f})(\beta,\mu_{\infty};\cdot\,)$ is identically zero.

Getting back to the susceptibility formula in \eqref{chiC} and by
virtue of the previous remarks, we expect to obtain an expansion of
the orbital susceptibility of the type \eqref{exformula2}.
The next two results identify the \textit{functions $\mathfrak{c}_{j_{1},l}(\cdot\,)$} coming from \eqref{trace1'} and \eqref{trace2'}:

\begin{lema}
\label{result1}
The quantity defined by \eqref{trace1'} can be rewritten as:
\begin{equation}
\label{retrace1'}
\mathrm{Tr}_{L^{2}(\mathbb{R}^{3})}\big\{\chi_{\Omega}
\mathcal{W}_{\infty,1}(\beta,\mu_{\infty})\big\} = - \frac{1}{4}
\frac{1}{\vert \Omega^{*}\vert} \sum_{j_{1} = 1}^{\infty}
\int_{\Omega^{*}} \mathrm{d}\mathbf{k}\,  \sum_{l = 0}^{3}
\frac{\partial^{l} \mathfrak{f}}{\partial \xi^{l}}
\big(\beta,\mu_{\infty};E_{j_{1}}(\mathbf{k})\big) \mathfrak{a}_{j_{1},l}(\mathbf{k})
\end{equation}
where for all $j_{1} \in \mathbb{N}^{*}$ and $\mathbf{k} \in \Omega^{*}$, the functions $\mathfrak{a}_{j_{1},3}(\cdot)$ and
$\mathfrak{a}_{j_{1},2}(\cdot)$ are given by:
\begin{multline}
\label{a3}
\mathfrak{a}_{j_{1},3}(\mathbf{k}) :=
\frac{1}{3!}\bigg\{ \big\vert
\hat{\pi}_{j_{1},j_{1}}(1;\mathbf{k})\big\vert^{2}
\sum_{\substack{j_{2} = 1 \\ j_{2} \neq j_{1}}}^{\infty}
\frac{\big\vert
\hat{\pi}_{j_{1},j_{2}}(2;\mathbf{k}) \big\vert^{2}}{E_{j_{2}}(\mathbf{k}) - E_{j_{1}}(\mathbf{k})} +
\big\vert \hat{\pi}_{j_{1},j_{1}}(2;\mathbf{k})\big\vert^{2}
\sum_{\substack{j_{2} = 1 \\ j_{2} \neq j_{1}}}^{\infty}
\frac{\big\vert \hat{\pi}_{j_{1},j_{2}}(1;\mathbf{k})
\big\vert^{2}}{E_{j_{2}}(\mathbf{k}) - E_{j_{1}}(\mathbf{k})}  \\
-
\hat{\pi}_{j_{1},j_{1}}(1;\mathbf{k})\hat{\pi}_{j_{1},j_{1}}(2;\mathbf{k})
\sum_{\substack{j_{2} = 1 \\ j_{2} \neq j_{1}}}^{\infty}  \frac{2
  \Re\big(\hat{\pi}_{j_{1},j_{2}}(2;\mathbf{k})
\hat{\pi}_{j_{2},j_{1}}(1;\mathbf{k})\big)}{E_{j_{2}}(\mathbf{k}) - E_{j_{1}}(\mathbf{k})}\bigg\}
\end{multline}
and
\begin{multline}
\label{a2}
\mathfrak{a}_{j_{1},2}(\mathbf{k}) := -\frac{1}{2!}\bigg\{  \sum_{\substack{j_{2} = 1 \\ j_{2} \neq j_{1}}}^{\infty}  \sum_{\substack{j_{3} = 1 \\ j_{3} \neq j_{1}}}^{\infty}  \frac{\mathcal{C}_{j_{1},j_{1},j_{2},j_{3}}(\mathbf{k}) + \mathcal{C}_{j_{1},j_{2},j_{1},j_{3}}(\mathbf{k}) + \mathcal{C}_{j_{1},j_{2},j_{3},j_{1}}(\mathbf{k})}{\big(E_{j_{2}}(\mathbf{k}) - E_{j_{1}}(\mathbf{k})\big)\big(E_{j_{3}}(\mathbf{k}) - E_{j_{1}}(\mathbf{k})\big)} \\
+ \sum_{\substack{j_{2} = 1 \\ j_{2} \neq j_{1}}}^{\infty}  \frac{\mathcal{C}_{j_{2},j_{1},j_{1},j_{1}}(\mathbf{k}) - \mathcal{C}_{j_{1},j_{1},j_{2},j_{1}}(\mathbf{k})}{\big(E_{j_{2}}(\mathbf{k}) - E_{j_{1}}(\mathbf{k})\big)^{2}}\bigg\}.
\end{multline}
\end{lema}
Note that it is possible to identify in \eqref{retrace1'} all the
functions $\mathfrak{a}_{j_{1},l}(\cdot)$ for $j_{1} \geq 1$ and
$l \in \{1,0\}$ since such a result is based only on identities
provided by the residue theorem. However, the number of terms is large
and we will not need their explicit expressions in order to prove our theorem.

\vspace{0.5cm}

Now we treat the next term.
\begin{lema}
\label{result2}
The quantity defined by \eqref{trace2'} can be rewritten as:
\begin{equation}
\label{retrace2'}
\mathrm{Tr}_{L^{2}(\mathbb{R}^{3})} \big\{\chi_{\Omega} \mathcal{W}_{\infty,2}(\beta,\mu_{\infty})\big\}= \frac{1}{4} \frac{1}{\vert \Omega^{*} \vert} \sum_{j_{1} =1}^{\infty} \int_{\Omega^{*}} \mathrm{d}\mathbf{k}\,  \sum_{l = 0}^{3} \frac{\partial^{l} \mathfrak{f}}{\partial \xi^{l}}\big(\beta,\mu_{\infty};E_{j_{1}}(\mathbf{k})\big) \mathfrak{b}_{j_{1},l}(\mathbf{k})
\end{equation}
where for all integers $j_{1} \geq 1$ and all $\mathbf{k} \in
\Omega^{*}$ we have:
\begin{gather}
\label{b3}
\mathfrak{b}_{j_{1},3}(\mathbf{k}) := \frac{1}{6} \big\{ \big\vert \hat{\pi}_{j_{1},j_{1}}(1;\mathbf{k})\big\vert^{2} + \big\vert \hat{\pi}_{j_{1},j_{1}}(2;\mathbf{k})\big\vert^{2}\big\}, \\
\label{b2}
\mathfrak{b}_{j_{1},2}(\mathbf{k}) := -\frac{1}{2}\sum_{\substack{j_{2} = 1 \\j_{2} \neq j_{1}}}^{\infty} \frac{\big\vert \hat{\pi}_{j_{1},j_{2}}(1;\mathbf{k})\big\vert^{2} + \big\vert \hat{\pi}_{j_{1},j_{2}}(2;\mathbf{k})\big\vert^{2}}{E_{j_{2}}(\mathbf{k}) - E_{j_{1}}(\mathbf{k})} +\frac{1}{2},\\
\mathfrak{b}_{j_{1},s}(\mathbf{k}) :=  -(2-s) \sum_{\substack{j_{2} = 1 \\j_{2} \neq j_{1}}}^{\infty} \frac{\big\vert \hat{\pi}_{j_{1},j_{2}}(1;\mathbf{k})\big\vert^{2} + \big\vert \hat{\pi}_{j_{1},j_{2}}(2;\mathbf{k})\big\vert^{2}}{\big(E_{j_{2}}(\mathbf{k}) - E_{j_{1}}(\mathbf{k})\big)^{3-s}},\quad s \in \{0,1\}. \nonumber
\end{gather}
\end{lema}

\vspace{0.5cm}

Thus our Lemmas \ref{result1} and \ref{result2} provide an expansion
of the type announced in \eqref{exformula2}, where the coefficients
are given by:
\begin{equation}
\label{coeff}
\mathfrak{c}_{j_{1},l}(\mathbf{k}) := \mathfrak{a}_{j_{1},l}(\mathbf{k}) + \mathfrak{b}_{j_{1},l}(\mathbf{k}),\quad l \in \{0,1,2,3\}.
\end{equation}
In particular, for all integer $j_{1} \geq 1$ and for all $\mathbf{k} \in \Omega^{*}$, the functions $\mathfrak{c}_{j_{1},3}(\cdot\,)$ and $\mathfrak{c}_{j_{1},2}(\cdot\,)$ are respectively given by:
\begin{multline}
\label{c3}
\mathfrak{c}_{j_{1},3}(\mathbf{k}) := \frac{1}{3!}\bigg\{\vert \hat{\pi}_{j_{1},j_{1}}(1;\mathbf{k})\vert^{2} \bigg(1+\sum_{\substack{j_{2} = 1 \\ j_{2} \neq j_{1}}}^{\infty} \frac{\vert
\hat{\pi}_{j_{1},j_{2}}(2;\mathbf{k}) \vert^{2}}{E_{j_{2}}(\mathbf{k}) - E_{j_{1}}(\mathbf{k})}\bigg) + \\
+ \vert \hat{\pi}_{j_{1},j_{1}}(2;\mathbf{k})\vert^{2} \bigg( 1 + \sum_{\substack{j_{2} = 1 \\ j_{2} \neq j_{1}}}^{\infty}\frac{\vert \hat{\pi}_{j_{1},j_{2}}(1;\mathbf{k})\vert^{2}}{E_{j_{2}}(\mathbf{k}) - E_{j_{1}}(\mathbf{k})}\bigg) +   \\
-\hat{\pi}_{j_{1},j_{1}}(1;\mathbf{k})\hat{\pi}_{j_{1},j_{1}}(2;\mathbf{k}) \sum_{\substack{j_{2} = 1 \\ j_{2} \neq j_{1}}}^{\infty}  \frac{2\Re\big(\hat{\pi}_{j_{1},j_{2}}(2;\mathbf{k}) \hat{\pi}_{j_{2},j_{1}}(1;\mathbf{k})\big)}{E_{j_{2}}(\mathbf{k}) - E_{j_{1}}(\mathbf{k})}\bigg\}
\end{multline}
and
\begin{multline}
\label{c2}
\mathfrak{c}_{j_{1},2}(\mathbf{k}) := -\frac{1}{2}\bigg\{ \sum_{\substack{j_{2} = 1 \\ j_{2} \neq j_{1}}}^{\infty} \frac{\vert \hat{\pi}_{j_{1},j_{2}}(1;\mathbf{k})\vert^{2} + \vert \hat{\pi}_{j_{1},j_{2}}(2;\mathbf{k})\vert^{2}}{E_{j_{2}}(\mathbf{k}) - E_{j_{1}}(\mathbf{k})} - 1 + \sum_{\substack{j_{2} = 1 \\ j_{2} \neq j_{1}}}^{\infty}  \frac{\mathcal{C}_{j_{2},j_{1},j_{1},j_{1}}(\mathbf{k}) - \mathcal{C}_{j_{1},j_{1},j_{2},j_{1}}(\mathbf{k})}{(E_{j_{2}}(\mathbf{k}) - E_{j_{1}}(\mathbf{k}))^{2}} + \\
+ \sum_{\substack{j_{2} = 1 \\ j_{2} \neq j_{1}}}^{\infty}  \sum_{\substack{j_{3} = 1 \\ j_{3} \neq j_{1}}}^{\infty}  \frac{\mathcal{C}_{j_{1},j_{1},j_{2},j_{3}}(\mathbf{k}) + \mathcal{C}_{j_{1},j_{2},j_{1},j_{3}}(\mathbf{k}) + \mathcal{C}_{j_{1},j_{2},j_{3},j_{1}}(\mathbf{k})}{(E_{j_{2}}(\mathbf{k}) - E_{j_{1}}(\mathbf{k}))(E_{j_{3}}(\mathbf{k}) - E_{j_{1}}(\mathbf{k}))}\bigg\},
\end{multline}
where for all integers $j_{1},j_{2},j_{3},j_{4} \in \mathbb{N}^{*}$, $\Omega^{*} \owns \mathbf{k} \mapsto \mathcal{C}_{j_{1},j_{2},j_{3},j_{4}}(\mathbf{k})$ is defined in \eqref{cj4}.
\vspace{0.5cm}

In order to conclude the proof of Theorem \ref{genexpand}, it remains to use this last result (its proof is in the appendix of this section):
\begin{lema}
\label{regulcof}
For all integers $j_{1} \geq 1$ and $l \in \{0,1,2,3\}$, the maps
$\Omega^{*} \owns \mathbf{k} \mapsto \mathfrak{a}_{j_{1},l}(\mathbf{k})$
and $\Omega^{*} \owns \mathbf{k} \mapsto
\mathfrak{b}_{j_{1},l}(\mathbf{k})$ are bounded and continuous on any
compact subset of $\Omega^*$ where $E_{j_1}$ is isolated from the rest
of the spectrum.
\end{lema}

Thus for all integers $j_{1} \geq 1$ and $\mathbf{k} \in \Omega^{*}$, the maps
$\mathfrak{c}_{j_{1},l}(\cdot\,)$ appearing in \eqref{exformula2} might be
singular on a set with zero Lebesgue measure where $E_{j_1}$
can touch the neighboring bands. However, the whole integrand in
\eqref{exformula2} is bounded and continuous on the whole $\Omega^*$
because it comes from two complex integrals (\eqref{trace1'} and
\eqref{trace2'}) which do not have local singularities in $\mathbf{k}$.

\subsection{Appendix - Proofs of the intermediate results}

Here we prove Lemmas \ref{result1}, \ref{result2}, and \ref{regulcof}.

\noindent \textbf{\textit{Proof of Lemma \ref{result1}}}.
Let $\Omega^{*} \owns \mathbf{k} \mapsto
\mathcal{C}_{j_{1},j_{2},j_{3},j_{4}}(\mathbf{k})$ be
the complex-valued function appearing in \eqref{trace1'}:
\begin{multline}\label{proszto1}
\mathcal{C}_{j_{1},j_{2},j_{3},j_{4}}(\mathbf{k}) := \big\{ \hat{\pi}_{j_{1},j_{2}}(1;\mathbf{k}) \hat{\pi}_{j_{2},j_{3}}(2;\mathbf{k}) - \hat{\pi}_{j_{1},j_{2}}(2;\mathbf{k})\hat{\pi}_{j_{2},j_{3}}(1;\mathbf{k})\big\} \\
\times \big\{ \hat{\pi}_{j_{3},j_{4}}(2;\mathbf{k}) \hat{\pi}_{j_{4},j_{1}}(1;\mathbf{k}) - \hat{\pi}_{j_{3},j_{4}}(1;\mathbf{k})\hat{\pi}_{j_{4},j_{1}}(2;\mathbf{k})\big\}.
\end{multline}
Note that this function is identically zero for the following
combinations of subscripts:
\begin{equation}
\label{combin}
j_{1} = j_{2} = j_{3} = j_{4},\quad j_{1} = j_{2} = j_{3} \neq j_{4},\quad j_{1} = j_{3} = j_{4} \neq j_{2}.
\end{equation}
Therefore the expansion of \eqref{trace1'} consists of partial derivatives of  $\mathfrak{f}(\beta,\mu_{\infty};\cdot\,)$ of order at most equal to three.
On the other hand, since the functions $\mathcal{C}_{j_{1},j_{1},j_{1},j_{4}}(\cdot\,)$ and $\mathcal{C}_{j_{1},j_{2},j_{1},j_{1}}(\cdot\,)$ are identically equal to zero (see \eqref{combin}), the quadruple summation in \eqref{trace1'} is reduced to :
\begin{multline}
\label{eq2'}
\sum_{j_{1},\ldots,j_{4}=1}^{\infty} \mathcal{C}_{j_{1},j_{2},j_{3},j_{4}}(\mathbf{k})  \bigg(\frac{1}{2i\pi}\bigg) \int_{\Gamma} \mathrm{d}\xi\, \frac{\mathfrak{f}(\beta,\mu_{\infty};\xi)}{\big(E_{j_{1}}(\mathbf{k}) - \xi\big)^{2} \big(E_{j_{2}}(\mathbf{k}) - \xi\big) \big(E_{j_{3}}(\mathbf{k}) - \xi\big) \big(E_{j_{4}}(\mathbf{k}) - \xi\big)} \\
= \sum_{j_{1} =1}^{\infty} \sum_{\substack{j_{3} = 1 \\ j_{3} \neq j_{1}}}^{\infty} \mathcal{C}_{j_{1},j_{1},j_{3},j_{1}}(\mathbf{k}) \bigg(\frac{1}{2i\pi}\bigg) \int_{\Gamma} \mathrm{d}\xi\, \frac{\mathfrak{f}(\beta,\mu_{\infty};\xi)}{\big(E_{j_{1}}(\mathbf{k}) - \xi\big)^{4} \big(E_{j_{3}}(\mathbf{k}) - \xi\big)}  \\
+ \sum_{j_{1} =1}^{\infty} \sum_{\substack{j_{2} = 1 \\ j_{2} \neq j_{1}}}^{\infty} \mathcal{C}_{j_{1},j_{2},j_{2},j_{2}}(\mathbf{k}) \bigg(\frac{1}{2i\pi}\bigg) \int_{\Gamma} \mathrm{d}\xi\, \frac{\mathfrak{f}(\beta,\mu_{\infty};\xi)}{\big(E_{j_{1}}(\mathbf{k}) - \xi\big)^{2} \big(E_{j_{2}}(\mathbf{k}) - \xi\big)^{3}}  \\
+\underbrace{\sum_{j_{1},\ldots,j_{4}=1}^{\infty}}_{\substack{\textrm{at most 2 equal} \\ \textrm{subscripts}}} \mathcal{C}_{j_{1},j_{2},j_{3},j_{4}}(\mathbf{k}) \bigg(\frac{1}{2i\pi}\bigg) \int_{\Gamma} \mathrm{d}\xi\, \frac{\mathfrak{f}(\beta,\mu_{\infty};\xi)}{\big(E_{j_{1}}(\mathbf{k}) - \xi\big)^{2} \big(E_{j_{2}}(\mathbf{k}) - \xi\big) \big(E_{j_{3}}(\mathbf{k}) - \xi\big) \big(E_{j_{4}}(\mathbf{k}) - \xi\big)}.
\end{multline}
By applying the residue theorem in the first term of the right hand
side of \eqref{eq2'} we get:
\begin{multline}
\label{eq3'}
\sum_{\substack{j_{3} = 1 \\ j_{3} \neq j_{1}}}^{\infty} \mathcal{C}_{j_{1},j_{1},j_{3},j_{1}}(\mathbf{k}) \bigg(\frac{1}{2i\pi}\bigg) \int_{\Gamma} \mathrm{d}\xi\, \frac{\mathfrak{f}(\beta,\mu_{\infty};\xi)}{\big(E_{j_{1}}(\mathbf{k}) - \xi\big)^{4} \big(E_{j_{3}}(\mathbf{k}) - \xi\big)} =
\sum_{\substack{j_{3} = 1 \\ j_{3} \neq j_{1}}}^{\infty} \mathcal{C}_{j_{1},j_{1},j_{3},j_{1}}(\mathbf{k}) \\  \bigg\{ \frac{1}{3!} \frac{1}{E_{j_{3}}(\mathbf{k}) - E_{j_{1}}(\mathbf{k})}  \frac{\partial^{3} \mathfrak{f}}{\partial \xi^{3}}\big(\beta,\mu_{\infty};E_{j_{1}}(\mathbf{k})\big) + \frac{3}{3!} \frac{1}{\big(E_{j_{3}}(\mathbf{k}) - E_{j_{1}}(\mathbf{k})\big)^{2}} \frac{\partial^{2} \mathfrak{f}}{\partial \xi^{2}}\big(\beta,\mu_{\infty};E_{j_{1}}(\mathbf{k})\big) +\\
+ \textrm{others terms involving $\frac{\partial^{l} \mathfrak{f}}{\partial \xi^{l}}(\beta,\mu_{\infty};\cdot)$, with $l \leq 1$} \bigg\}.
\end{multline}
The function $\mathcal{C}_{j_{1},j_{1},j_{3},j_{1}}(\cdot\,)$ appearing in front of $\frac{\partial^{3} \mathfrak{f}}{\partial \xi^{3}}\big(\beta,\mu_{\infty};E_{j_{1}}(\mathbf{k})\big)$ in \eqref{eq3'} corresponds to $\mathfrak{a}_{j_{1},3}(\cdot\,)$ since:
\begin{equation*}
\forall\, \mathbf{k} \in \Omega^{*}, \quad \mathcal{C}_{j_{1},j_{1},j_{3},j_{1}}(\mathbf{k}) = \big\vert \hat{\pi}_{j_{1},j_{1}}(1;\mathbf{k}) \hat{\pi}_{j_{1},j_{3}}(2;\mathbf{k}) - \hat{\pi}_{j_{1},j_{1}}(2;\mathbf{k}) \hat{\pi}_{j_{1},j_{3}}(1;\mathbf{k}) \big\vert^{2}.
\end{equation*}
Note that the function
$\mathcal{C}_{j_{1},j_{1},j_{3},j_{1}}(\cdot)$ contributes to the
term $\mathfrak{a}_{j_{1},2}(\cdot)$, too.

By applying once again the residue theorem in the second term of the
right hand side of \eqref{eq2'} we obtain:
\begin{multline}
\label{eq4'}
\sum_{\substack{j_{2} = 1 \\ j_{2} \neq j_{1}}}^{\infty} \mathcal{C}_{j_{1},j_{2},j_{2},j_{2}}(\mathbf{k}) \bigg(\frac{1}{2i\pi}\bigg) \int_{\Gamma} \mathrm{d}\xi\, \frac{\mathfrak{f}(\beta,\mu_{\infty};\xi)}{\big(E_{j_{1}}(\mathbf{k}) - \xi\big)^{2} \big(E_{j_{3}}(\mathbf{k}) - \xi\big)^{3}}  \\
=\sum_{\substack{j_{2} = 1 \\ j_{2} \neq j_{1}}}^{\infty} \mathcal{C}_{j_{1},j_{2},j_{2},j_{2}}(\mathbf{k}) \bigg\{ - \frac{1}{2!} \frac{1}{\big(E_{j_{1}}(\mathbf{k}) - E_{j_{2}}(\mathbf{k})\big)^{2}}  \frac{\partial^{2} \mathfrak{f}}{\partial \xi^{2}}\big(\beta,\mu_{\infty};E_{j_{2}}(\mathbf{k})\big)\\
+ \textrm{others terms involving $\frac{\partial^{l} \mathfrak{f}}{\partial \xi^{l}}(\beta,\mu_{\infty};\cdot)$, with $l \leq 1$} \bigg\}.
\end{multline}
The function $\mathcal{C}_{j_{1},j_{2},j_{2},j_{2}}(\cdot\,)$
appearing in front of $\frac{\partial^{2} \mathfrak{f}}{\partial
  \xi^{2}}\big(\beta,\mu_{\infty};E_{j_{2}}(\mathbf{k})\big)$
contributes to $\mathfrak{a}_{j_{1},2}(\cdot\,)$.

It remains to isolate in \eqref{eq2'}
(where at most two subscripts are equal) all combinations
which provide a second order derivative of $\mathfrak{f}(\beta,\mu_{\infty};\cdot\,)$. These combinations are:
\begin{equation*}
j_{1} = j_{2} \neq j_{3}, j_{4};\quad j_{1} = j_{3} \neq j_{2},j_{4};\quad j_{1} = j_{4} \neq j_{2}, j_{3}.
\end{equation*}
Finally, we once again apply the residue theorem and gathering all
terms proportional with $\frac{\partial^{2} \mathfrak{f}}{\partial
  \xi^{2}}(\beta,\mu_{\infty};\cdot\,)$. The proof is over.
  \qed

\noindent \textbf{\textit{Proof of Lemma \ref{result2}}}. By separating the cases $j_{1} = j_{2}$ and $j_{1} \neq j_{2}$,
the double summation in the right hand side of \eqref{trace2'} reads as:
\begin{multline}
\label{eq1'}
\sum_{j_{1}=1}^{\infty} \sum_{j_{2} = 1}^{\infty} \mathcal{C}_{j_{1},j_{2}}(\mathbf{k}) \bigg(\frac{1}{2i\pi}\bigg) \int_{\Gamma} \mathrm{d}\xi\, \frac{\mathfrak{f}(\beta,\mu_{\infty};\xi)}{\big(E_{j_{1}}(\mathbf{k}) - \xi\big)^{3} \big(E_{j_{2}}(\mathbf{k}) - \xi\big)} \\
=  \sum_{j_{1} = 1}^{\infty} \mathcal{C}_{j_{1},j_{1}}(\mathbf{k})
 \bigg(\frac{1}{2i\pi}\bigg) \int_{\Gamma} \mathrm{d}\xi\, \frac{\mathfrak{f}(\beta,\mu_{\infty};\xi)}{\big(E_{j_{1}}(\mathbf{k}) - \xi\big)^{4}}  \\
+ \sum_{j_{1} = 1}^{\infty} \sum_{\substack{j_{2} =1 \\ j_{2} \neq j_{1}}}^{\infty} \mathcal{C}_{j_{1},j_{2}}(\mathbf{k}) \bigg(\frac{1}{2i\pi}\bigg) \int_{\Gamma} \mathrm{d}\xi\, \frac{\mathfrak{f}(\beta,\mu_{\infty};\xi)}{\big(E_{j_{1}}(k)(\mathbf{k}) - \xi\big)^{3} \big(E_{j_{2}}(\mathbf{k}) - \xi\big)}.
\end{multline}
By using the residue theorem in the first term of the r.h.s. of \eqref{eq1'} :
\begin{equation*}
\bigg(\frac{1}{2i\pi}\bigg) \int_{\Gamma} \mathrm{d}\xi\, \frac{\mathfrak{f}(\beta,\mu_{\infty};\xi)}{\big(E_{j_{1}}(\mathbf{k}) - \xi\big)^{4}} = \frac{1}{3!} \frac{\partial^{3} \mathfrak{f}}{\partial \xi^{3}}\big(\beta,\mu_{\infty};E_{j_{1}}(\mathbf{k})\big)
\end{equation*}
This is only the one term which provides a third-order partial
derivative of $\mathfrak{f}(\beta,\mu_{\infty};\cdot\,)$. The rest of
the proof is just plain computation using the residue theorem. We do
not give further details. \qed

\noindent \textbf{\textit{Proof of Lemma \ref{regulcof}}}.
Let $p_{\alpha} := -i\partial_{\alpha}$
be the $\alpha$ component of the momentum operator with periodic
boundary conditions in $L^2(\Omega)$, $\alpha \in \{1,2,3\}$.
Now assume that $E_{j_1}(\mathbf{k})$ is isolated and non-degenerate if $\mathbf{k}$ belongs to some
compact $K\subset \Omega^*$. We have to investigate integrals of the type
\begin{align}\label{martie01-11}
 {\rm Tr}_{L^2(\Omega)}\int_\Gamma \mathrm{d}\xi\,\mathfrak{f}(\beta,\mu_{\infty};\xi)
(h(\mathbf{k})-\xi)^{-1}p_{\alpha_1}(h(\mathbf{k})-\xi)^{-1}\dotsb p_{\alpha_4}(h(\mathbf{k})-\xi)^{-1}.
\end{align}
Let $\mathbf{k}_0\in K$, and let $\Gamma_1$ be a simple, positively oriented path surrounding $E_{j_1}(\mathbf{k}_0)$ but no other eigenvalue of
$h(\mathbf{k}_0)$. If $|\mathbf{k}-\mathbf{k}_0|$ is small enough, $\Gamma_1$ will still only contain $E_{j_1}(\mathbf{k})$.
The projection $\Pi(\mathbf{k})$ corresponding to $E_{j_1}(\mathbf{k})$ is given by a Riesz integral. We have:
\begin{align}\label{martie02-11}
 \Pi(\mathbf{k})=\frac{i}{2\pi}\int_{\Gamma_1} \mathrm{d}z\, (h(\mathbf{k})-z)^{-1},
\end{align}
and is continuous at $\mathbf{k}_0$ in the trace norm topology. Moreover,
\begin{align}\label{martie03-11}
\Pi(\mathbf{k})(h(\mathbf{k})-\xi)^{-1}=\frac{1}{E_{j_1}(\mathbf{k})-\xi}\Pi(\mathbf{k}),\; (\mathbf {1}-\Pi(\mathbf{k}))(h(\mathbf{k})-\xi)^{-1}=
\frac{1}{2\pi i}\int_{\Gamma_1} \mathrm{d}z\,\frac{1}{z-\xi} (h(\mathbf{k})-z)^{-1}.
\end{align}
Clearly, $\Pi(\mathbf{k})(h(\mathbf{k})-\xi)^{-1}$ is analytic in $\xi$ in the exterior of $\Gamma_1$. We can decompose the integral on $\Gamma$ in
\eqref{martie01-11} as a sum of three integrals, one of which being on a simple contour $\Gamma_2$ around $E_{j_1}(\mathbf{k}_0)$, completely surrounded by
$\Gamma_1$. The other two integrals will never have
$E_{j_1}(\mathbf{k})$ as a singularity, so they cannot contribute to the formula of $\mathfrak{a}_{j_{1},l}(\mathbf{k})$. On the other hand,
in the integral on $\Gamma_2$ we can replace the resolvents with the decomposition in  \eqref{martie03-11} and use the fact that
$(\mathbf{1}-\Pi(\mathbf{k}))(h(\mathbf{k})-\xi)^{-1}$ is analytic if $\xi$ lies inside $\Gamma_2$. Now one can apply the Cauchy residue formula. For example,
we can compute the integral in which we have $\Pi(\mathbf{k})$ at the
extremities, and $(\mathbf{1}-\Pi(\mathbf{k}))$ in the interior; in that case
$E_{j_1}=E_{j_1}(\mathbf{k})$ will be a double pole:
\begin{align}\label{martie04-11}
 &{\rm Tr}_{L^2(\Omega)}\int_{\Gamma_2} \mathrm{d}\xi\,\mathfrak{f}(\xi)
\Pi(\mathbf{k})(h(\mathbf{k})-\xi)^{-1}p_{\alpha_1}(h(\mathbf{k})-\xi)^{-1}(\mathbf{1}-\Pi(\mathbf{k}))
\dotsb p_{\alpha_4}(h(\mathbf{k})-\xi)^{-1}\Pi(\mathbf{k})
\nonumber \\
&=2\pi i \bigg\{(\partial_{\xi}\mathfrak{f})(E_{j_1}(\mathbf{k})){\rm Tr}_{L^2(\Omega)}\left\{ \Pi(\mathbf{k})p_{\alpha_1}(h(\mathbf{k})-E_{j_1})^{-1}(\mathbf{1}-\Pi(\mathbf{k}))
\dotsb p_{\alpha_4}\Pi(\mathbf{k})\right\} + \\
&+ \mathfrak{f}(E_{j_1}(\bold{k}))\frac{d}{d\xi}{\rm Tr}_{L^2(\Omega)}\left \{\Pi(\mathbf{k})p_{\alpha_1}
(h(\mathbf{k})-\xi)^{-1}(\mathbf{1}-\Pi(\mathbf{k}))
\dotsb (h(\mathbf{k})-\xi)^{-1}(\mathbf{1}-\Pi(\mathbf{k}))p_{\alpha_4}\right\}_{\xi=E_{j_1}(\mathbf{k})}\bigg\}.\nonumber
\end{align}
Thus one contribution to $\mathfrak{a}_{j_{1},1}(\mathbf{k})$ will be:
$${\rm Tr}_{L^2(\Omega)}\left\{ \Pi(\mathbf{k})p_{\alpha_1}(h(\mathbf{k})-E_{j_1})^{-1}(\mathbf{1}-\Pi(\mathbf{k}))
\dotsb p_{\alpha_4}\Pi(\mathbf{k})\right\}.$$
This expression does not use eigenvectors, only resolvents and projectors. Since $E_{j_1}$ is continuous at $\mathbf{k}_0$, the map
$$\mathbf{k}\mapsto (\mathbf{1}-\Pi(\mathbf{k}))(h(\mathbf{k})-E_{j_1}(\mathbf{k}))^{-1}=
\frac{1}{2\pi i}\int_{\Gamma_1} \mathrm{d}z\, \frac{1}{z-E_{j_1}(\mathbf{k})} (h(\mathbf{k})-z)^{-1}$$
is operator norm continuous at $\mathbf{k}_0$, and the map
$\mathbf{k}\mapsto \Pi(\mathbf{k})$ is continuous in the trace norm.
By using standard perturbation theory (see e.g. \cite{K}), the same holds for the maps:
$$ \mathbf{k}\mapsto (\mathbf{1}-\Pi(\mathbf{k}))(h(\mathbf{k})-E_{j_1}(\mathbf{k}))^{-1}p_{\alpha_{l}}
\quad { \rm and } \quad \mathbf{k}\mapsto p_{\alpha_{l}}(\mathbf{1}-\Pi(\mathbf{k}))(h(\mathbf{k})-E_{j_1}(\mathbf{k}))^{-1}p_{\alpha_{k}}.$$
Thus the trace defines a
continuous function; all other coefficients can be treated in a similar way.
\qed

\section{The zero-field susceptibility at fixed density and zero temperature}

In this section, we separately investigate the
semiconducting and metallic cases from the expansion \eqref{exformula2}. In particular,
we prove  Theorem \ref{maintheorem} $(i)$ and $(ii)$.

\subsection{The semiconducting case (SC)- Proof of Theorem \ref{maintheorem} $(i)$}

By using that $\mathfrak{f}_{FD}(\beta,\mu;\xi) = - \beta^{-1} \partial_{\xi} \mathfrak{f}(\beta,\mu;\xi)$, \eqref{exformula2} can be rewritten as:
\begin{multline}
\label{exformula3}
\mathcal{X}(\beta,\rho_{0})=\bigg(\frac{e}{c}\bigg)^{2} \frac{1}{2}
\frac{1}{(2\pi)^3} \\
 \sum_{j_{1}=1}^{\infty} \int_{\Omega^{*}} \mathrm{d}\mathbf{k}\,
\bigg\{\sum_{l= 0}^{2} \frac{\partial^{l} \mathfrak{f}_{FD}}{\partial
  \xi^{l}}\big(\beta,\mu_{\infty};E_{j_{1}}(\mathbf{k})\big)
\mathfrak{c}_{j_{1},1+l}(\mathbf{k}) - \frac{1}{\beta} \mathfrak{f}\big(\beta,\mu_{\infty};E_{j_{1}}(\mathbf{k})\big) \mathfrak{c}_{j_{1},0}(\mathbf{k})\bigg\}.
\end{multline}
From \eqref{exformula3}, the proof of Theorem \ref{maintheorem}  $(i)$ is based on two
main ingredients. The first one is that for any fixed $\mu \geq E_0$ we have the following
pointwise convergences:
\begin{equation}
\label{pointwise}
\lim_{\beta \rightarrow \infty} \frac{1}{\beta}
\mathfrak{f}(\beta,\mu;\xi) = (\mu - \xi)
\chi_{[E_{0},\mu]}(\xi),\quad \lim_{\beta \rightarrow \infty}
\mathfrak{f}_{FD}(\beta,\mu;\xi) = \chi_{[E_{0},\mu]}(\xi),\quad
\forall\, \xi \in [E_{0},\infty)\setminus \{\mu\},\quad
\end{equation}
while in the distributional sense:
\begin{equation}
\label{distributional}
\lim_{\beta \rightarrow  \infty} \frac{\partial
  \mathfrak{f}_{FD}}{\partial \xi}(\beta,\mu;\xi) = - \delta(\xi -
\mu),\quad \lim_{\beta \rightarrow  \infty} \frac{\partial^{2}
  \mathfrak{f}_{FD}}{\partial \xi^{2}}(\beta,\mu;\xi) = -\partial_\xi \delta(\xi -
\mu).
\end{equation}
The second ingredient is related to the decay of the derivatives
of the Fermi-Dirac distribution: for all $d > 0$ and for all
$j \in \mathbb{N}^{*}$, there exists a constant $C_{j,d} > 0$ such that
\begin{equation}
\label{prosty1}
\sup_{\vert\xi-\mu\vert\geq d>0}\bigg \vert \frac{\partial^j
  \mathfrak{f}_{FD}}{\partial \xi^j}(\beta,\mu;\xi)\bigg\vert\leq C_{j,d}
\mathrm{e}^{- \frac{\beta \vert \xi-\mu\vert}{2}}.
\end{equation}

Now assume that we are in the semiconducting case with a non-trivial
gap, that is there exists
$N \in \mathbb{N}^{*}$ such that
$\lim_{\beta \rightarrow  \infty} \mu_{\infty}(\beta,\rho_{0}) =
(\max \mathcal{E}_{N} + \min \mathcal{E}_{N+1})/2 =
\mathcal{E}_{F}(\rho_{0})$ and $\max \mathcal{E}_{N}<\min
\mathcal{E}_{N+1}$. Since the Fermi energy lies inside a gap, all
terms containing derivatives of the Fermi-Dirac distribution will
converge to zero in the limit
$\beta \rightarrow \infty$. Here \eqref{prosty1} plays a double important
role: first, it makes the series in $j_1$ convergent, and second, it
provides an exponential decay to zero. Then by taking into account
\eqref{pointwise},
we immediately get \eqref{susemicon} from \eqref{exformula3}.
\qed

\subsection{The metallic case (M)- Proof of Theorem \ref{maintheorem} $(ii)$}

Now we are interested in the metallic case. The limit $\beta\to\infty$
is not so simple as in the previous case, because the Fermi energy
lies in the spectrum. The starting point is the same formula \eqref{exformula2}, but
 we have to modify it by getting rid of the third order partial derivatives of
$\mathfrak{f}$ in order to make appear a
Landau-Peierls type contribution. However, this operation needs the already announced
additional assumption of
non-degeneracy (which will provide regularity in $\mathbf{k}$) in a neighborhood of
the Fermi surface:
\begin{assumption}
\label{assumption}
We assume that there exists a
unique $N \in \mathbb{N}^{*}$ such that
$\lim_{\beta \rightarrow  \infty} \mu_{\infty}(\beta,\rho_{0})
=\mathcal{E}_F(\rho_{0})\in (\min \mathcal{E}_{N},\max \mathcal{E}_{N})$,
which means that the Fermi energy lies inside
the $N$th Bloch band
$\mathcal{E}_{N}$. We also  assume that the Fermi surface defined by
$\mathcal{S}_{F} := \{\mathbf{k} \in \Omega^{*}\,:\,E_{N}(\mathbf{k}) = \mathcal{E}_{F}(\rho_{0})\}$
is smooth and non-degenerate.
\end{assumption}
Recall that $E_N(\mathbf{k})$ is supposed to be non degenerate outside a (possibly empty) zero Lebesgue
measure set of $\mathbf{k}$-points.
\vspace{0.5cm}
Our assumption leads to the following consequence:
\begin{equation}\label{prostye1}
{\rm dist}\big\{\mathcal{E}_{F}(\rho_{0}), \cup_{j=1}^{N-1}
\mathcal{E}_{j}\big\} = d_{1} > 0,\quad {\rm
  dist}\big\{\mathcal{E}_{F}(\rho_{0}),
\cup_{j=N+1}^{\infty} \mathcal{E}_{j}\big\}= d_{2} > 0.
\end{equation}
Note that the minimum of the lowest Bloch band $\mathcal{E}_{1}$ is always simple.
If the density $\rho_0$ is small enough then Assumption \ref{assumption} is automatically
satisfied since the Bloch energy function $\mathbf{k} \mapsto E_{1}(\mathbf{k})$ is
non-degenerate in a neighborhood of $\mathbf{k}=\mathbf{0}$ (see e.g. \cite{KS}).

In fact, the non-degeneracy assumption is indispensable for to use
of the regular perturbation theory in order to express the functions defined by
\eqref{b3}, \eqref{b2} and \eqref{a3} (only in the case where $j_{1} =
N$) with the help of the partial derivatives of $E_{N}(\cdot)$ with respect to
the $k_{i}$-variables, for $\mathbf{k}$ in a neighborhood of the Fermi surface:
\begin{align}
\label{der1}
\frac{\partial E_{N}(\mathbf{k})}{\partial k_{i}} &=
\hat{\pi}_{N,N}(i;\mathbf{k}),\quad i\in\{1,2,3\}, \\
\label{der2}
\frac{\partial^{2} E_{N}(\mathbf{k})}{\partial k_{i}^{2}} &= 1 + 2 \sum_{\substack{j = 1 \\ j \neq N}}^{\infty} \frac{\big\vert \hat{\pi}_{j,N}(i;\mathbf{k}) \big\vert^{2}}{E_{N}(\mathbf{k}) - E_{j}(\mathbf{k})},\quad i\in\{1,2,3\}, \\
\label{der2croi}
\frac{\partial^{2} E_{N}(\mathbf{k})}{\partial k_{1} \partial k_{2}} &= \sum_{\substack{j = 1 \\ j \neq N}}^{\infty} \frac{2 \Re \big\{ \hat{\pi}_{j,N}(1;\mathbf{k})\hat{\pi}_{N,j}(2;\mathbf{k})\big\}}{E_{N}(\mathbf{k}) - E_{j}(\mathbf{k})} = \frac{\partial^{2} E_{N}(\mathbf{k})}{\partial k_{2} \partial k_{1}}.
\end{align}
Such identities have been studied in \cite{CN}.
Note that the above series are absolutely convergent if
the potential $V$
is smooth enough (\cite{CN}).

Now using Assumption \ref{assumption}, we can group the coefficients corresponding to
the third and
second order derivatives of $\mathfrak{f}$ appearing in \eqref{exformula2}.
This operation allows us to isolate a
Landau-Peierls type contribution (the proof can be found in the appendix of
this section):
\begin{proposition}
\label{propo3}
Assume for simplicity that $\mathcal{E}_{N}$ is a simple band. Let $\Omega^{*} \owns \mathbf{k}
\mapsto \mathfrak{c}_{N,2}(\mathbf{k})$ and $\Omega^{*} \owns \mathbf{k}
\mapsto \mathfrak{c}_{N,3}(\mathbf{k})$ the functions respectively defined by \eqref{c2} and \eqref{c3} with $j_{1}= N$.
Then:
\begin{multline}
\label{expant}
 \int_{\Omega^{*}} \mathrm{d}\mathbf{k}\, \sum_{l=2}^{3} \frac{\partial^{l} \mathfrak{f}}{\partial \xi^{l}}\big(\beta,\mu_{\infty}; E_{N}(\mathbf{k})\big)\mathfrak{c}_{N,l}(\mathbf{k}) \\
= \int_{\Omega^{*}} \mathrm{d}\mathbf{k}\, \frac{\partial^{2} \mathfrak{f}}{\partial \xi^{2}}\big(\beta,\mu_{\infty}; E_{N}(\mathbf{k})\big) \bigg\{\frac{1}{3!} \frac{\partial^{2}E_{N}(\mathbf{k})}{\partial k_{1}^{2}}\; \frac{\partial^{2}E_{N}(\mathbf{k})}{\partial k_{2}^{2}} - \frac{1}{3!}\bigg(\frac{\partial^{2}E_{N}(\mathbf{k})}{\partial k_{1}\partial k_{2}}\bigg)^{2} + \mathfrak{a}_{N,2}(\mathbf{k}) \bigg\},
\end{multline}
where $\Omega^{*} \owns \mathbf{k} \mapsto \mathfrak{a}_{j_{1},2}(\mathbf{k})$ are
the functions defined in \eqref{a2}.
\end{proposition}

\vspace{0.5cm}

From \eqref{exformula2} and Proposition \ref{propo3} we get an
expansion for the orbital susceptibility at fixed density
$\rho_{0}>0$ and inverse of temperature $\beta > 0$:
\begin{proposition}
\label{globform}
Assume for simplicity that $\mathcal{E}_{N}$ is a simple band. For every
$j_{1} \in \mathbb{N}^{*}$ there exist four families of
functions $\mathfrak{c}_{j_{1},l}(\cdot)$
with $l \in \{0,1,2,3\}$, defined on $\Omega^{*}$ outside a set of Lebesgue measure zero,
such that the second integrand below is bounded and continuous on $\Omega^*$:
\begin{multline}
\label{mainformula}
\mathcal{X}(\beta,\rho_{0}) = -\bigg(\frac{e}{c}\bigg)^{2}\frac{1}{12\beta} \frac{1}{(2\pi)^3} \cdot \\
\bigg\{\int_{\Omega^{*}} \mathrm{d}\mathbf{k}\, \frac{\partial^{2} \mathfrak{f}}{\partial \xi^{2}}\big(\beta,\mu_{\infty};E_{N}(\mathbf{k})\big)
\bigg[\frac{\partial^{2}E_{N}(\mathbf{k})}{\partial k_{1}^{2}} \frac{\partial^{2}E_{N}(\mathbf{k})}{\partial k_{2}^{2}} -
\bigg(\frac{\partial^{2}E_{N}(\mathbf{k})}{\partial k_{1}\partial k_{2}}\bigg)^{2} - 3 \mathcal{F}_{N}(\mathbf{k}) \bigg]  \\
+ 6 \int_{\Omega^{*}} \mathrm{d}\mathbf{k}\, \bigg[\sum_{\substack{j_{1} = 1 \\ j_{1} \neq N}}^{\infty}  \sum_{l =2}^{3}  \frac{\partial^{l} \mathfrak{f}}{\partial\xi^{l}} \big(\beta,\mu_{\infty};E_{j_{1}}(\mathbf{k})\big) \mathfrak{c}_{j_{1},l}(\mathbf{k}) + \sum_{j_{1} = 1}^{\infty}  \sum_{l = 0}^{1}  \frac{\partial^{l} \mathfrak{f}}{\partial\xi^{l}} \big(\beta,\mu_{\infty};E_{j_{1}}(\mathbf{k})\big) \mathfrak{c}_{j_{1},l}(\mathbf{k})\bigg] \bigg\},
\end{multline}
where by convention $(\partial_{\xi}^{0} \mathfrak{f})(\beta,\mu_{\infty};\cdot) := \mathfrak{f}(\beta,\mu_{\infty};\cdot)$ and:
\begin{multline}
\label{FN}
\mathcal{F}_{N}(\mathbf{k}) := - 2 \mathfrak{a}_{N,2}(\bold{k}) = \sum_{\substack{j_{2} = 1 \\ j_{2} \neq N}}^{\infty} \sum_{\substack{j_{3} =
1 \\ j_{3} \neq N}}^{\infty} \frac{\mathcal{C}_{N,N,j_{2},j_{3}}(\mathbf{k}) + \mathcal{C}_{N,j_{2},N,j_{3}}(\mathbf{k}) +
\mathcal{C}_{N,j_{2},j_{3},N}(\mathbf{k})}{\big(E_{j_{2}}(\mathbf{k}) - E_{N}(\mathbf{k})\big)\big(E_{j_{3}}(\mathbf{k}) - E_{N}(\mathbf{k})\big)} \\
+ \sum_{\substack{j_{2} = 1 \\ j_{2} \neq N}}^{\infty} \frac{\mathcal{C}_{j_{2},N,N,N}(\mathbf{k}) - \mathcal{C}_{N,N,j_{2},N}(\mathbf{k})}{\big(E_{j_{2}}(\mathbf{k}) - E_{N}(\mathbf{k})\big)^{2}}.
\end{multline}
\end{proposition}
\vspace{0.5cm}
Note that we can use identities provided by the regular perturbation theory in order to express the functions $\mathfrak{c}_{j_1,l}$ (as well as $\mathcal{F}_{N}$) appearing in \eqref{mainformula} in terms of derivatives of $E_{j}$ and $u_{j}$ w.r.t. the ${\bf k}$-variable. But this formulation will only hold true outside a set of ${\bf k}$-points of Lebesgue measure zero, while the formulation involving
$\hat{\pi}_{i,j}$'s is more general, physically relevant, providing us with bounded and continuous coefficients
on $\Omega^*$ (see Lemma \ref{regulcof}).
Finally keep in mind that the main goal is the Landau-Peierls formula, and it will turn out that only
the factor multiplying the second partial derivative of $\mathfrak{f}$ will contribute to it.

In order to complete the proof of Theorem \ref{maintheorem} $(ii)$, it remains to take
the limit when
$\beta \rightarrow \infty$ in \eqref{mainformula}. Since the Fermi
energy lies inside the band $\mathcal{E}_{N}$ and it is isolated from
all other bands, then using \eqref{distributional} and \eqref{prosty1} we have:
\begin{equation*}
\lim_{\beta \rightarrow  \infty} \frac{1}{\beta}\int_{\Omega^{*}} \mathrm{d}\mathbf{k}\, \sum_{\substack{j=1 \\ j \neq N}}^{\infty} \sum_{l=2}^{3} \frac{\partial^{l} \mathfrak{f}}{\partial \xi^{l}}(\beta,\mu_{\infty}(\beta,\rho_0);E_{j}(\mathbf{k}))\mathfrak{c}_{j,l}(\mathbf{k}) = 0
\end{equation*}
and
\begin{multline*}
\lim_{\beta \rightarrow  \infty} - \frac{1}{\beta} \int_{\Omega^{*}} \mathrm{d}\mathbf{k}\,  \frac{\partial^{2} \mathfrak{f}}{\partial \xi^{2}}(\beta,\mu_{\infty};E_{N}(\mathbf{k})) \bigg\{\frac{\partial^{2}E_{N}(\mathbf{k})}{\partial k_{1}^{2}} \frac{\partial^{2}E_{N}(\mathbf{k})}{\partial k_{2}^{2}} - \bigg(\frac{\partial^{2}E_{N}(\mathbf{k})}{\partial k_{1}\partial k_{2}}\bigg)^{2} - 3 \mathcal{F}_{N}(\mathbf{k}) \bigg\} \\
= -\int_{\mathcal{S}_{F}} \frac{\mathrm{d}\sigma(\mathbf{k})}{\vert \nabla E_{N}(\mathbf{k}) \vert} \bigg\{\frac{\partial^{2}E_{N}(\mathbf{k})}{\partial k_{1}^{2}} \frac{\partial^{2}E_{N}(\mathbf{k})}{\partial k_{2}^{2}} - \bigg(\frac{\partial^{2}E_{N}(\mathbf{k})}{\partial k_{1}\partial k_{2}}\bigg)^{2} - 3 \mathcal{F}_{N}(\mathbf{k}) \bigg\}
\end{multline*}
where $\mathcal{S}_{F}$ denotes the Fermi surface. Using these two
identities together with \eqref{pointwise} in \eqref{mainformula}, we obtain \eqref{sumetal}.

\subsection{Appendix - Proof of Proposition \ref{propo3}}

Using \eqref{coeff} we get:
\begin{multline*}
\int_{\Omega^{*}} \mathrm{d}\mathbf{k}\, \sum_{l=2}^{3} \frac{\partial^{l} \mathfrak{f}}{\partial \xi^{l}}\big(\beta,\mu_{\infty}; E_{N}(\mathbf{k})\big)\mathfrak{c}_{N,l}(\mathbf{k}) =
\int_{\Omega^{*}} \mathrm{d}\mathbf{k}\, \frac{\partial^{2} \mathfrak{f}}{\partial \xi^{2}}\big(\beta,\mu_{\infty}; E_{N}(\mathbf{k})\big) \mathfrak{a}_{N,2}(\mathbf{k})  \\
+
\int_{\Omega^{*}} \mathrm{d}\mathbf{k}\,\bigg[ \sum_{l=2}^{3} \frac{\partial^{l} \mathfrak{f}}{\partial \xi^{l}}\big(\beta,\mu_{\infty}; E_{N}(\mathbf{k})\big) \mathfrak{b}_{N,l}(\mathbf{k}) + \frac{\partial^{3} \mathfrak{f}}{\partial \xi^{3}}\big(\beta,\mu_{\infty}; E_{N}(\mathbf{k})\big)  \mathfrak{a}_{N,3}(\mathbf{k})\bigg].
\end{multline*}
Using \eqref{der1} and \eqref{der2}, the functions $\mathfrak{b}_{N,l}(\cdot)$, $l \in \{2,3\}$, can be rewritten as:
\begin{align*}
\mathfrak{b}_{N,3}(\mathbf{k}) &= \frac{1}{3!} \bigg\{\bigg(\frac{\partial E_{N}(\mathbf{k})}{\partial k_{1}}\bigg)^{2} + \bigg(\frac{\partial E_{N}(\mathbf{k})}{\partial k_{2}}\bigg)^{2}\bigg\}, \\
\mathfrak{b}_{N,2}(\mathbf{k}) &= -\frac{1}{2!}
\bigg\{-\frac{1}{2}\bigg(\frac{\partial^{2} E_{N}(\mathbf{k})}{\partial
  k_{1}^{2}} - 1\bigg) - \frac{1}{2}\bigg(\frac{\partial^{2}
  E_{N}(\mathbf{k})}{\partial k_{2}^{2}} - 1\bigg) - 1\bigg\} \\
&= \frac{1}{2!}\frac{1}{2} \bigg\{\frac{\partial^{2} E_{N}(\mathbf{k})}{\partial k_{1}^{2}} + \frac{\partial^{2} E_{N}(\mathbf{k})}{\partial k_{2}^{2}}\bigg\}.
\end{align*}
Since $E_{N}(\cdot\,) \in \mathcal{C}^{2}(\mathbb{R}^{3}/(2 \pi \mathbb{Z}^{3}))$, a simple integration by parts gives us:
\begin{multline}
\label{eq5'}
\forall\, i \in \{1,2\},\quad \int_{-\pi}^{\pi} \mathrm{d}k_{i}\, \frac{\partial E_{N}(\mathbf{k})}{\partial k_{i}} \frac{\partial \mathfrak{f}^{3}}{\partial \xi^{3}}\big(\beta,\mu_{\infty};E_{N}(\mathbf{k})\big) \frac{\partial E_{N}(\mathbf{k})}{\partial k_{i}} \\
= - \int_{-\pi}^{\pi} \mathrm{d}k_{i}\, \frac{\partial \mathfrak{f}^{2}}{\partial \xi^{2}}\big(\beta,\mu_{\infty};E_{N}(\mathbf{k})\big) \frac{\partial^{2} E_{N}(\mathbf{k})}{\partial k_{i}^{2}}
\end{multline}
whence:
\begin{equation*}
\int_{\Omega^{*}} \mathrm{d}\mathbf{k}\, \frac{\partial^{3} \mathfrak{f}}{\partial \xi^{3}}\big(\beta,\mu_{\infty}; E_{N}(\mathbf{k})\big)\mathfrak{b}_{N,3}(\mathbf{k}) = - \frac{1}{3!} \int_{\Omega^{*}} \mathrm{d}\mathbf{k}\, \frac{\partial^{2} \mathfrak{f}}{\partial \xi^{2}}\big(\beta,\mu_{\infty}; E_{N}(\mathbf{k})\big)\bigg\{\frac{\partial^{2} E_{N}(\mathbf{k})}{\partial k_{1}^{2}} + \frac{\partial^{2} E_{N}(\mathbf{k})}{\partial k_{2}^{2}}\bigg\}
\end{equation*}
and:
\begin{equation}
\label{eq6'}
\int_{\Omega^{*}} \mathrm{d}\mathbf{k}\, \sum_{l = 2}^{3} \frac{\partial^{l} \mathfrak{f}}{\partial \xi^{l}}\big(\beta,\mu_{\infty}; E_{N}(\mathbf{k})\big)\mathfrak{b}_{N,l}(\mathbf{k}) = \frac{1}{3!} \frac{1}{2} \int_{\Omega^{*}} \mathrm{d}\mathbf{k}\, \frac{\partial^{2} \mathfrak{f}}{\partial \xi^{2}}\big(\beta,\mu_{\infty}; E_{N}(\mathbf{k})\big)\bigg\{\frac{\partial^{2} E_{N}(\mathbf{k})}{\partial k_{1}^{2}} + \frac{\partial^{2} E_{N}(\mathbf{k})}{\partial k_{2}^{2}}\bigg\}
\end{equation}
On the other hand, using \eqref{der1}, \eqref{der2} and \eqref{der2croi}, the function $\mathfrak{a}_{N,3}(\cdot\,)$ can be rewritten as :
\begin{multline}
\label{eq7'}
\mathfrak{a}_{N,3}(\mathbf{k}) = \frac{1}{3!} \bigg\{\bigg(\frac{\partial E_{N}(\mathbf{k})}{\partial k_{1}}\bigg)^{2} \frac{1}{2} \bigg(1 - \frac{\partial^{2} E_{N}(\mathbf{k})}{\partial k_{2}^{2}}\bigg) + \bigg(\frac{\partial E_{N}(\mathbf{k})}{\partial k_{2}}\bigg)^{2} \frac{1}{2} \bigg(1 - \frac{\partial^{2} E_{N}(\mathbf{k})}{\partial k_{1}^{2}}\bigg) \\
- \bigg(\frac{\partial E_{N}(\mathbf{k})}{\partial k_{1}}\bigg)\bigg(\frac{\partial E_{N}(\mathbf{k})}{\partial k_{2}}\bigg) \bigg(-\frac{\partial^{2} E_{N}(\mathbf{k})}{\partial k_{1} \partial k_{2}}\bigg)\bigg\}.
\end{multline}
Note that by a simple integration by parts:
\begin{align}
\label{eq8'}
&\forall\, i\neq j \in\{1,2\},\quad \int_{-\pi}^{\pi} \mathrm{d}k_{j}\, \frac{\partial E_{N}(\mathbf{k})}{\partial k_{j}}{} \frac{\partial^{3} \mathfrak{f}}{\partial \xi^{3}}\big(\beta,\mu_{\infty};E_{N}(\mathbf{k})\big){} \frac{\partial E_{N}(\mathbf{k})}{\partial k_{j}}{}\frac{\partial^{2} E_{N}(\mathbf{k})}{\partial k_{i}^{2}} \\
&= - \int_{-\pi}^{\pi} \mathrm{d}k_{j}\, \frac{\partial^{2} \mathfrak{f}}{\partial \xi^{2}}\big(\beta,\mu_{\infty};E_{N}(\mathbf{k})\big){} \frac{\partial}{\partial k_{j}} \bigg[\frac{\partial E_{N}(\mathbf{k})}{\partial k_{j}}{}\frac{\partial^{2} E_{N}(\mathbf{k})}{\partial k_{i}^{2}}\bigg] \nonumber \\
&= - \int_{-\pi}^{\pi} \mathrm{d}k_{j}\, \frac{\partial^{2} \mathfrak{f}}{\partial \xi^{2}}\big(\beta,\mu_{\infty};E_{N}(\mathbf{k})\big){} \bigg\{\frac{\partial^{2} E_{N}(\mathbf{k})}{\partial k_{j}^{2}}{} \frac{\partial^{2}E_{N}(\mathbf{k})}{\partial k_{i}^{2}} + \frac{\partial E_{N}(\mathbf{k})}{\partial k_{j}} \frac{\partial}{\partial k_{j}} \frac{\partial^{2} E_{N}(\mathbf{k})}{\partial k_{i}^{2}}\bigg\}\nonumber \\
&= - \int_{-\pi}^{\pi} \mathrm{d}k_{j}\, \frac{\partial^{2} \mathfrak{f}}{\partial \xi^{2}}\big(\beta,\mu_{\infty};E_{N}(\mathbf{k})\big){} \bigg\{\frac{\partial^{2}E_{N}(\mathbf{k})}{\partial k_{j}^{2}} {} \frac{\partial^{2} E_{N}(\mathbf{k})}{\partial k_{i}^{2}} + \frac{\partial E_{N}(\mathbf{k})}{\partial k_{j}} \frac{\partial}{\partial k_{i}} \frac{\partial^{2} E_{N}(\mathbf{k})}{\partial k_{j}\partial k_{i}}\bigg\}\nonumber.
\end{align}
By virtue of \eqref{eq7'}, using \eqref{eq8'} and \eqref{eq5'}, we get:
\begin{multline}
\label{eq9'}
\int_{\Omega^{*}} \mathrm{d}\mathbf{k}\, \frac{\partial^{3} \mathfrak{f}}{\partial \xi^{3}}\big(\beta,\mu_{\infty};E_{N}(\mathbf{k})\big){}\mathfrak{a}_{N,3}(\mathbf{k}) = \frac{1}{3!} \frac{1}{2} \int_{\Omega^{*}} \mathrm{d}\mathbf{k}\, \frac{\partial^{2} \mathfrak{f}}{\partial \xi^{2}}\big(\beta,\mu_{\infty};E_{N}(\mathbf{k})\big){} \bigg\{2 \frac{\partial^{2} E_{N}(\mathbf{k})}{\partial k_{1}^{2}} {} \frac{\partial^{2} E_{N}(\mathbf{k})}{\partial k_{2}^{2}} \\
+ \frac{\partial E_{N}(\mathbf{k})}{\partial k_{1}} \frac{\partial}{\partial k_{2}} \frac{\partial^{2} E_{N}(\mathbf{k})}{\partial k_{1}\partial k_{2}} + \frac{\partial E_{N}(\mathbf{k})}{\partial k_{2}} \frac{\partial}{\partial k_{1}} \frac{\partial^{2} E_{N}(\mathbf{k})}{\partial k_{2}\partial k_{1}} - \frac{\partial^{2} E_{N}(\mathbf{k})}{\partial k_{1}^{2}} - \frac{\partial^{2} E_{N}(\mathbf{k})}{\partial k_{2}^{2}} \bigg\}  \\
+\frac{1}{3!} \int_{\Omega^{*}} \mathrm{d}\mathbf{k}\, \frac{\partial^{3} \mathfrak{f}}{\partial \xi^{3}}\big(\beta,\mu_{\infty};E_{N}(\mathbf{k})\big){} \frac{\partial E_{N}(\mathbf{k})}{\partial k_{1}}{} \frac{\partial E_{N}(\mathbf{k})}{\partial k_{2}} {} \frac{\partial^{2} E_{N}(\mathbf{k})}{\partial k_{1} \partial k_{2}}.
\end{multline}
Finally, by a last integration by parts:
\begin{align*}
&\forall\, i\neq j \in\{1,2\},\quad \int_{-\pi}^{\pi} \mathrm{d}k_{j}\, \frac{\partial E_{N}(\mathbf{k})}{\partial k_{j}}{} \frac{\partial^{2} \mathfrak{f}}{\partial \xi^{2}}\big(\beta,\mu_{\infty};E_{N}(\mathbf{k})\big){} \frac{\partial}{\partial k_{i}}\frac{\partial^{2} E_{N}(\mathbf{k})}{\partial k_{j} \partial k_{i}}\\
&= - \int_{-\pi}^{\pi} \mathrm{d}k_{j}\, \frac{\partial}{\partial k_{i}} \bigg[\frac{\partial E_{N}(\mathbf{k})}{\partial k_{j}}{} \frac{\partial^{2} \mathfrak{f}}{\partial \xi^{2}}\big(\beta,\mu_{\infty};E_{N}(\mathbf{k})\big)\bigg]{} \frac{\partial^{2} E_{N}(\mathbf{k})}{\partial k_{j} \partial k_{i}} \\
&= - \int_{-\pi}^{\pi} \mathrm{d}k_{j}\, \bigg\{\frac{\partial^{2}E_{N}(\mathbf{k})}{\partial k_{i} \partial k_{j}}{} \frac{\partial^{2} \mathfrak{f}}{\partial \xi^{2}}\big(\beta,\mu_{\infty};E_{N}(\mathbf{k})\big) +  \frac{\partial E_{N}(\mathbf{k})}{\partial k_{j}}{}\frac{\partial E_{N}(\mathbf{k})}{\partial k_{i}} {} \frac{\partial^{3} \mathfrak{f}}{\partial \xi^{3}}\big(\beta,\mu_{\infty};E_{N}(\mathbf{k})\big)\bigg\} {} \frac{\partial^{2}E_{N}(\mathbf{k})}{\partial k_{j} \partial k_{i}}.
\end{align*}
Then \eqref{eq9'} is reduced to:
\begin{multline}
\label{eq10'}
\int_{\Omega^{*}} \mathrm{d}\mathbf{k}\, \frac{\partial^{3} \mathfrak{f}}{\partial \xi^{3}}\big(\beta,\mu_{\infty};E_{N}(\mathbf{k})\big){}\mathfrak{a}_{N,3}(\mathbf{k})
= \frac{1}{3!}\frac{1}{2} \int_{\Omega^{*}} \mathrm{d}\mathbf{k}\, \frac{\partial^{2} \mathfrak{f}}{\partial \xi^{2}}\big(\beta,\mu_{\infty};E_{N}(\mathbf{k})\big){} \\
{} \bigg\{2\frac{\partial^{2} E_{N}(\mathbf{k})}{\partial k_{1}^{2}}{} \frac{\partial^{2} E_{N}(\mathbf{k})}{\partial k_{2}^{2}} - 2 \bigg(\frac{\partial^{2} E_{N}(\mathbf{k})}{\partial k_{1} \partial k_{2}}\bigg)^{2}
-  \frac{\partial^{2} E_{N}(\mathbf{k})}{\partial k_{1}^{2}} - \frac{\partial^{2} E_{N}(\mathbf{k})}{\partial k_{2}^{2}}\bigg\}.
\end{multline}
By adding \eqref{eq6'} to \eqref{eq10'} we get :
\begin{multline*}
\int_{\Omega^{*}} \mathrm{d}\mathbf{k}\, \bigg[ \sum_{l=2}^{3} \frac{\partial^{l} \mathfrak{f}}{\partial \xi^{l}}\big(\beta,\mu_{\infty}; E_{N}(\mathbf{k})\big) {} \mathfrak{b}_{N,l}(\mathbf{k}) + \frac{\partial^{3} \mathfrak{f}}{\partial \xi^{3}}\big(\beta,\mu_{\infty}; E_{N}(\mathbf{k})\big) {} \mathfrak{a}_{N,3}(\mathbf{k})\bigg] \\
= \frac{1}{3!} \int_{\Omega^{*}} \mathrm{d}\mathbf{k}\, \frac{\partial^{2} \mathfrak{f}}{\partial \xi^{2}}\big(\beta,\mu_{\infty}; E_{N}(\mathbf{k})\big) \bigg\{ \frac{\partial^{2}E_{N}(\mathbf{k})}{\partial k_{1}^{2}}{} \frac{\partial^{2}E_{N}(\mathbf{k})}{\partial k_{2}^{2}} - \bigg(\frac{\partial^{2}E_{N}(\mathbf{k})}{\partial k_{1}\partial k_{2}}\bigg)^{2}\bigg\}
\end{multline*}
and we are done. Note that the proof does not work if $E_N$ can touch
other bands because we loose regularity. In that case the integration
by parts have to be done across a tubular neighborhood of the Fermi
surface $\mathcal{S}_F$, the price being the apparition of some extra
terms. These terms will though disappear in the limit $\beta\to
\infty$ because they will decay exponentially with $\beta$.
\qed
\section{The Landau-Peierls formula}

The aim of this section is to establish an asymptotic expansion of
\eqref{sumetal} in the limit of small densities ($\rho_{0} \rightarrow
0$). Here we prove the expansion \eqref{sulp}, of which \eqref{chilp} is a particular case which has allready been suggested by T. Kjeldaas and W. Kohn in 1957 \cite{KK}.

\subsection{Proof of Theorem \ref{maintheorem}  $(iii)$}
Let us recall that $E_0=\min_{\mathbf{k} \in \Omega^{*}}
E_{1}(\mathbf{k})=E_1(\mathbf{0})$, and $E_{1}(\mathbf{k})$ is non degenerate
near the origin with a positive definite Hessian matrix (see e.g. \cite{KS}).
The same reference insures the existence of the
following quadratic expansion of $E_{1}(\mathbf{k})$ for $\mathbf{k}
\rightarrow \mathbf{0}$:
\begin{equation*}
E_{1}(\mathbf{k}) = E_{0} + \frac{1}{2!}\mathbf{k}^{T}\Bigg[\frac{\partial^{2} E_{1}}{\partial k_{i} \partial k_{j}}(\mathbf{0})\Bigg]_{1 \leq i,j\leq 3} \mathbf{k} + \mathcal{O}\big(\mathbf{k}^{4}\big)\quad \textrm{when\,\,$\mathbf{k} \rightarrow 0$}
\end{equation*}
As the Hessian matrix is symmetric, then up to a change of coordinates
this quadratic expansion can be rewritten as :
\begin{equation}
\label{expqua}
E_{1}(\mathbf{k}) = E_{0} + \frac{1}{2} \sum_{i = 1}^{3} \frac{k_{i}^{2}}{m_{i}^{*}} + \mathcal{O}\big(\mathbf{k}^{4}\big)\quad \textrm{when\,\,$\mathbf{k} \rightarrow 0$}
\end{equation}
where $\big[\frac{1}{m_{i}^{*}}\big]_{1\leq i \leq 3}$ are the
eigenvalues of the inverse effective-mass tensor.

Consider the assumption of weak density $\rho_{0} \in
(0,1)$. In this case the Fermi energy defined by \eqref{ferenergy}
lies in the interval $(E_0,\max_{\mathbf{k} \in \Omega^{*}}
E_{1}(\mathbf{k}))$. When $\rho_{0} \rightarrow 0$ it follows that
$\mathcal{E}_{F}(\rho_{0})$ converges to $E_0$. The $\mathbf{k}$-subset
of $\Omega^*$
where $E_0\leq E_1(\mathbf{k})\leq \mathcal{E}_{F}(\rho_{0})$ is therefore
only localized near the origin.

From \eqref{expqua}
we get the following asymptotic expansion of
$\mathcal{E}_{F}(\rho_{0}) - E_{0}$ when $\rho_{0} \rightarrow 0$ (the
proof is given in the appendix of this section):
\begin{proposition}
\label{propo4}
When $\rho_{0} \rightarrow 0$, we have the following expansion:
\begin{equation}
\label{expfermi}
\mathcal{E}_{F}(\rho_{0}) - E_{0} = s \rho_{0}^{\frac{2}{3}} + \mathcal{O}\big(\rho_{0}^{\frac{4}{3}}\big),\qquad s := \frac{(6\pi^{2})^{\frac{2}{3}}}{2} \bigg(\frac{1}{m_{1}^{*}m_{2}^{*}m_{3}^{*}}\bigg)^{\frac{1}{3}}.
\end{equation}
In the particular case when $m_{i}^{*} = m^{*}>0$ for $i \in \{1,2,3\}$ and by setting $k_{F} := (6\pi^{2} \rho_{0})^{\frac{1}{3}}$:
\begin{equation}
\label{expfermisp}
\mathcal{E}_{F}(\rho_{0}) - E_{0} = \frac{1}{2 m^{*}} k_{F}^{2} + \mathcal{O}\big(k_{F}^{4}\big).
\end{equation}
\end{proposition}

\vspace{0.5cm}

Before proving Theorem \ref{maintheorem} $(iii)$, we need one more technical result
(its proof is also in the appendix of this section):
\begin{lema}
\label{needing}
Assume that $E_1(\mathbf{k})$ remains non-degenerate on the ball $
B_{\epsilon_0}(\mathbf{0}):=\{\mathbf{k}\in \Omega^*\,:\, \vert\mathbf{k}\vert\leq
\epsilon_0\}$ with $\epsilon_0>0$ small enough. Consider any continuous function
$F : B_{\epsilon_0}(\mathbf{0}) \rightarrow \mathbb{C}$. Then when $\rho_{0} \rightarrow 0$ we
have the following asymptotic expansions:
\begin{align}
\label{first}
\int_{\mathcal{S}_{F}} \frac{\mathrm{d}\sigma(\mathbf{k})}{\big\vert \nabla E_{1}(\mathbf{k}) \big\vert}\,F(\mathbf{k}) = A \rho_{0}^{\frac{1}{3}} + o\big(\rho_{0}^{\frac{1}{3}}\big) \quad
{\rm with} \quad A:= \sqrt{m_{1}^{*} m_{2}^{*} m_{3}^{*}} 4\sqrt{2}\pi F(\mathbf{0}) \sqrt{s}
\end{align}
and
\begin{align}
\label{second}
\int_{\Omega^{*}} \mathrm{d}\mathbf{k}\,
\chi_{[E_{0},\mathcal{E}_{F}(\rho_{0})]}\big(E_{1}(\mathbf{k})\big)
F(\mathbf{k})  = B \rho_{0} + o\big(\rho_{0}\big)\quad
{\rm with} \quad B:= \sqrt{m_{1}^{*} m_{2}^{*} m_{3}^{*}} \frac{8 \sqrt{2} \pi}{3} F(\mathbf{0}) s^{\frac{3}{2}},
\end{align}
where $s$ is the coefficient defined in \eqref{expfermi}.
\end{lema}

\vspace{0.5cm}

Now we are ready to prove the Landau-Peierls formula in \eqref{sulp}. For this, consider
the formula \eqref{sumetal}. Remember that $E_{1}(\cdot)$ is non-degenerate and
analytic in a neighborhood of
the origin. Let us concentrate ourselves on the first term appearing in \eqref{sumetal}:
\begin{equation}
\label{termnu1}
- \bigg(\frac{e}{c}\bigg)^{2} \frac{1}{12} \frac{1}{(2\pi)^3}
\int_{\mathcal{S}_{F}} \frac{\mathrm{d}\sigma(\mathbf{k})}{\big\vert
  \nabla E_{1}(\mathbf{k})
  \big\vert}\,\bigg\{\frac{\partial^{2}E_{1}(\mathbf{k})}{\partial
  k_{1}^{2}}\; \frac{\partial^{2}E_{1}(\mathbf{k})}{\partial k_{2}^{2}}
- \bigg(\frac{\partial^{2}E_{1}(\mathbf{k})}{\partial k_{1}\partial
  k_{2}}\bigg)^{2} - 3 \mathcal{F}_{1}(\mathbf{k})\bigg\},
\end{equation}
since only this term will have a nonzero contribution to
the leading term in \eqref{sulp}. The other term will go to zero like
$\rho_0$; this can be shown using
\eqref{second}, \eqref{expqua}, and the fact that the coefficients
$\mathfrak{c}_{1,1}$ and $\mathfrak{c}_{1,0}$ are continuous near
$\mathbf{0}$ (see Lemma \ref{regulcof}).

Now consider the following function:
$$F(\mathbf{k}):=\frac{\partial^{2}E_{1}(\mathbf{k})}{\partial
  k_{1}^{2}}\; \frac{\partial^{2}E_{1}(\mathbf{k})}{\partial k_{2}^{2}}
- \bigg(\frac{\partial^{2}E_{1}(\mathbf{k})}{\partial k_{1}\partial
  k_{2}}\bigg)^{2} - 3 \mathcal{F}_{1}(\mathbf{k}).$$
By taking into account that $\mathcal{F}_{1}(\cdot\,) = -2 \mathfrak{a}_{1,2}(\cdot\,)$ (see \eqref{FN}) and by virtue of Lemma \ref{regulcof}, $F(\cdot)$ is continuous near the origin.
According to \eqref{first},
the only thing we need to do is to compute $F(\mathbf{0})$. The determinant of
the Hessian matrix gives after a short computation:
\begin{equation}
\label{fisident}
\frac{\partial^{2} E_{1}(\mathbf{k})}{\partial k_{1}^{2}} \frac{\partial^{2} E_{1}(\mathbf{k})}{\partial k_{2}^{2}} - \bigg(\frac{\partial^{2} E_{1}(\mathbf{k})}{\partial k_{1}\partial k_{2}}\bigg)^{2} = \frac{1}{m_{1}^{*}m_{2}^{*}} + \mathcal{O}\big(\mathbf{k}^{2}\big)\quad \textrm{when $\mathbf{k} \rightarrow \mathbf{0}$}.
\end{equation}
Thus we can write:
\begin{equation*}
\mathcal{X}_{\rm M}(\rho_{0})
= - \bigg(\frac{e}{c}\bigg)^{2} \frac{1}{24 \pi^{2}} \big(m_{1}^{*}m_{2}^{*}m_{3}^{*}\big)^{\frac{1}{3}} \bigg[\frac{1}{m_{1}^{*} m_{2}^{*}} - 3 \mathcal{F}_{1}(\mathbf{0})\bigg](6 \pi)^{\frac{1}{3}}\rho_{0}^{\frac{1}{3}} + o\big(\rho_{0}^{\frac{1}{3}}\big)\quad \textrm{when $\rho_{0} \rightarrow 0$}.
\end{equation*}
The only thing we have left to do, is proving that
$\mathcal{F}_{1}(\mathbf{0}) = 0$. The definition of $\mathcal{F}_{1}$
can be found in \eqref{FN}, while the coefficients entering in its definition are
defined in \eqref{proszto1}.

Let us start by showing that for all integers $j_{2}, j_{3} \geq 2$ we have:
\begin{equation*}
\mathcal{C}_{1,1,j_{2},j_{3}}(\mathbf{0}) =
\mathcal{C}_{1,j_{2},j_{3},1}(\mathbf{0}) =
\mathcal{C}_{j_{2},1,1,1}(\mathbf{0}) =
\mathcal{C}_{1,1,j_{2},1}(\mathbf{0}) = 0.
\end{equation*}
Indeed, in the expression of each of these functions it is possible to
identify a factor of the type
$\hat{\pi}_{1,1}(\alpha;\mathbf{0})$, $\alpha \in \{1,2\}$ which are
nothing but partial derivatives of $E_1$ at the origin, thus they must
be zero. It follows that:
\begin{equation}
\label{F1}
\mathcal{F}_{1}(\mathbf{0}) = \sum_{j_{2} = 2}^{\infty} \sum_{j_{3} = 2}^{\infty} \frac{\mathcal{C}_{1,j_{2},1,j_{3}}(\mathbf{0})}{\big(E_{j_{2}}(\mathbf{0}) - E_{1}(\mathbf{0})\big)\big(E_{j_{3}}(\mathbf{0}) - E_{1}(\mathbf{0})\big)}.
\end{equation}
Since:
\begin{multline*}
\mathcal{C}_{1,j_{2},1,j_{3}}(\mathbf{0}) =
\hat{\pi}_{1,j_{2}}(1;\mathbf{0})
\hat{\pi}_{j_{2},1}(2;\mathbf{0})\hat{\pi}_{1,j_{3}}(2;\mathbf{0})
\hat{\pi}_{j_{3},1}(1;\mathbf{0}) + \hat{\pi}_{1,j_{2}}(2;\mathbf{0})
\hat{\pi}_{j_{2},1}(1;\mathbf{0})\hat{\pi}_{1,j_{3}}(1;\mathbf{0})\hat{\pi}_{j_{3},1}(2;\mathbf{0})  \\
- \hat{\pi}_{1,j_{2}}(2;\mathbf{0})\hat{\pi}_{j_{2},1}(1;\mathbf{0}) \hat{\pi}_{1,j_{3}}(2;\mathbf{0})
\hat{\pi}_{j_{3},1}(1;\mathbf{0}) - \hat{\pi}_{1,j_{2}}(1;\mathbf{0})\hat{\pi}_{j_{2},1}(2;\mathbf{0}) \hat{\pi}_{1,j_{3}}(1;\mathbf{0})\hat{\pi}_{j_{3},1}(2;\mathbf{0}),
\end{multline*}
then \eqref{F1} can be rewritten as :
\begin{equation}
\label{F1'}
\mathcal{F}_{1}(\mathbf{0}) = 2 \bigg\vert \sum_{j=2}^{\infty}
\frac{\hat{\pi}_{1,j}(2;\mathbf{0})
  \hat{\pi}_{j,1}(1;\mathbf{0})}{E_{j}(\mathbf{0}) -
  E_{1}(\mathbf{0})}\bigg\vert^{2} -
\bigg(\sum_{j=2}^{\infty} \frac{\hat{\pi}_{1,j}(2;\mathbf{0}) \hat{\pi}_{j,1}(1;\mathbf{0})}{E_{j}(\mathbf{0}) - E_{1}(\mathbf{0})}\bigg)^{2} - \bigg(\overline{\sum_{j=2}^{\infty} \frac{\hat{\pi}_{1,j}(2;\mathbf{0}) \hat{\pi}_{j,1}(1;\mathbf{0})}{E_{j}(\mathbf{0}) - E_{1}(\mathbf{0})}}\bigg)^{2}.
\end{equation}
But for $\mathbf{k}=0$ we may choose all our eigenfunctions
$u_{l}(\cdot\,;\mathbf{0})$ to be real. It means that for
all integers $j \geq 2$ and $\alpha \in \{1,2\}$, the matrix elements
$\hat{\pi}_{1,j}(\alpha;\mathbf{0})$ are purely imaginary.
As a result, the sums in \eqref{F1'} are real numbers and cancel each
other, thus $\mathcal{F}_{1}(\mathbf{0}) = 0$.
\qed

\subsection{Appendix - Proofs of intermediate results}

Here we prove Proposition \ref{propo4} and Lemma \ref{needing}.

\noindent \textbf{\textit{Proof of Proposition \ref{propo4}}}. In \eqref{expqua} use the
change of variables $\tilde{k}_{i} :=
\frac{k_{i}}{\sqrt{m_{i}^{*}}}$, with $i \in \{1,2,3\}$. This gives:
\begin{equation*}
\tilde{E}_1(\tilde{\mathbf{k}}):=E_{1}(\sqrt{m_{i}^{*}}\;\tilde{\mathbf{k}}) = E_{0} + \frac{1}{2} \big\{\tilde{k}_{1}^2 +
\tilde{k}_{2}^2 + \tilde{k}_{3}^2 \big\} + \mathcal{O}(\tilde{\mathbf{k}}^{4}).
\end{equation*}
In spherical coordinates:
\begin{equation}
\label{4.1}
\tilde{E}_{1}(r,\theta,\phi) = E_{0} + \frac{1}{2} r^{2} +
\mathcal{O}(r^{4})\quad \textrm{when\,\,$r \rightarrow 0$}.
\end{equation}
We would like to express $r$ as a function of $\tilde{E}_{1}$,
$\theta$ and $\phi$. Clearly, the equation
$\tilde{E}_{1}\big(r(\theta,\phi),\theta,\phi\big) =E_0+\Delta$ has
a unique solution
$r(\theta,\phi,\Delta)$ if
$\Delta>0$ is small enough. This
solution obeys a fixed point equation of the type:
\begin{equation}
\label{4.2}
r(\theta,\phi,\Delta) = \sqrt{2 \Delta} \big[1 +
\mathcal{O}\big(r^{2}(\theta,\phi,\Delta)\big)\big]
\end{equation}
which leads to the estimate:
\begin{equation}
\label{4.3}
r(\theta,\phi,\Delta) = \sqrt{2 \Delta} \big[1 + \mathcal{O}\big(\Delta\big)\big]\quad \textrm{when\,\,$\Delta \rightarrow 0$}.
\end{equation}
We can finally determine $\Delta$ (thus the Fermi energy)
as a function $\rho_0$.
By setting $\tilde{\Omega}^{*} := \frac{\Omega^{*}}{\sqrt{m_{1}^{*} m_{2}^{*} m_{3}^{*}}}$, it follows from \eqref{limitrho}:
\begin{equation*}
\rho_{0} =
\frac{\sqrt{m_{1}^{*}m_{2}^{*}m_{3}^{*}}}{(2\pi)^3}
\int_{\tilde{\Omega}^{*}} \mathrm{d}\tilde{\mathbf{k}}\,
\chi_{[E_{0}, E_0+\Delta]}\big(\tilde{E}_{1}(\tilde{\mathbf{k}})\big).
\end{equation*}
Using spherical coordinates:
\begin{equation*}
\rho_{0} = \frac{\sqrt{m_{1}^{*}m_{2}^{*}m_{3}^{*}}}
{(2\pi)^3} \int_{0}^{2\pi} \mathrm{d}\phi\, \int_{0}^{\pi} \mathrm{d}\theta\, \sin \theta \bigg\{\int_{0}^{\sqrt{2 \Delta}} \mathrm{d}r\, r^{2} + \int_{\sqrt{2 \Delta}}^{r(\theta,\phi,\Delta)} \mathrm{d}r\, r^{2}\bigg\}.
\end{equation*}
This is the equation we have to solve in order to find $\Delta$ as a
function of $\rho_0$. Then by standard fixed point arguments we arrive
at the estimate \eqref{expfermi} and we are done.
\qed

\noindent \textbf{\textit{Proof of Lemma \ref{needing}}}. We only prove \eqref{first}, the other estimate being similar. As
before, we prefer the new variables $\tilde{k}_{i} =
\frac{k_{i}}{\sqrt{m_{i}^{*}}}$ where $i \in \{1,2,3\}$. Denote by
$\tilde{E}_1(\mathbf{\tilde{k}})=E_1(\mathbf{k})$, by
$\tilde{F}(\mathbf{\tilde{k}})=F(\mathbf{k})$ and with
$\tilde{\Omega}^{*} :=
\frac{\Omega^{*}}{\sqrt{m_{1}^{*}m_{2}^{*}m_{3}^{*}}}$. Then we can
formally write:
\begin{align}
\label{5.1}
\int_{\mathcal{S}_{F}} \frac{\mathrm{d}\sigma(\mathbf{k})}{\big\vert \nabla E_{1}(\mathbf{k}) \big\vert}\,F(\mathbf{k})& =\int_{\Omega^{*}} \mathrm{d}\mathbf{k}\, \delta\big(\mathcal{E}_{F}(\rho_{0}) - E_{1}(\mathbf{k})\big) F(\mathbf{k}) \nonumber\\
&= \sqrt{m_{1}^{*}m_{2}^{*}m_{3}^{*}} \int_{\tilde{\Omega}^{*}}
\mathrm{d}\mathbf{\tilde{k}}\, \delta\big(\mathcal{E}_{F}(\rho_{0}) -
\tilde{E}_{1}
(\tilde{\mathbf{k}})\big) \tilde{F}(\tilde{\mathbf{k}})
\nonumber \\
&= \sqrt{m_{1}^{*}m_{2}^{*}m_{3}^{*}} \int_{\{\,\tilde{\mathbf{k}}\in
  \tilde{\Omega}^{*}\,\,\textrm{s.t.}\,\tilde{E}_{1}(\tilde{\mathbf{k}})=
  \mathcal{E}_{F}(\rho_{0})\,\}}
\frac{\mathrm{d}\sigma(\mathbf{\tilde{k}})}{\vert
  \nabla_{\tilde{\mathbf{k}}} \tilde{E}_{1}(\tilde{\mathbf{k}}) \vert} \tilde{F}
(\tilde{\mathbf{k}}).
\end{align}
The quadratic expansion \eqref{expqua} implies $\vert
\nabla_{\tilde{\mathbf{k}}} \tilde{E}_{1}(\tilde{\mathbf{k}}) \vert =
\vert\tilde{\mathbf{k}}\vert\big[1 +
\mathcal{O}\big(\tilde{\mathbf{k}}^{2}\big)\big]$ when $\tilde{\mathbf{k}}
\rightarrow \mathbf{0}$.
Then:
\begin{align}
\label{5.2}\int_{\mathcal{S}_{F}} \frac{\mathrm{d}\sigma(\mathbf{k})}{\big\vert
  \nabla E_{1}(\mathbf{k}) \big\vert}\, F(\mathbf{k})
= \sqrt{m_{1}^{*} m_{2}^{*} m_{3}^{*}}F(\mathbf{0}) \int_{\{\,\tilde{\mathbf{k}} \in \tilde{\Omega}^{*}\,\,\textrm{s.t.}\,\,\tilde{E}_1(\tilde{\mathbf{k}}) = \mathcal{E}_{F}(\rho_{0})\,\}} \mathrm{d}\sigma(\tilde{\mathbf{k}})\,  \vert \tilde{\mathbf{k}}\vert^{-1}\big[1 + o(1)\big].
\end{align}
Using spherical coordinates, let us denote as before by
$r(\theta,\phi,\rho_0)$ the unique root of the equation
$\tilde{E}_{1}\big(r(\theta,\phi,\rho_0),\theta,\phi\big) =
\mathcal{E}_{F}(\rho_{0})$. Then \eqref{5.2} can be rewritten as:
\begin{align*}
\int_{\mathcal{S}_{F}} \frac{\mathrm{d}\sigma(\mathbf{k})}{\big\vert \nabla E_{1}(\mathbf{k}) \big\vert}\,F(\mathbf{k})
= \sqrt{m_{1}^{*}m_{2}^{*}m_{3}^{*}} F(\mathbf{0})\int_{0}^{2 \pi}\mathrm{d}\phi\, \int_{0}^{\pi} \mathrm{d}\theta\, \sin(\theta)\; r(\theta,\phi,\rho_0)\big[1 + o(1)\big].
\end{align*}
Now by setting $\Delta := \mathcal{E}_{F}(\rho_{0}) - E_{0}$ and by using \eqref{4.3} when $\Delta \rightarrow 0$:
\begin{equation*}
\int_{\mathcal{S}_{F}} \frac{\mathrm{d}\sigma(\mathbf{k})}{\big\vert \nabla E_{1}(\mathbf{k}) \big\vert}\,F(\mathbf{k}) =
\sqrt{m_{1}^{*} m_{2}^{*} m_{3}^{*}} \;4\sqrt{2}\pi F(\mathbf{0})\; \sqrt{\Delta}\big[1 + o(1)\big].
\end{equation*}
Finally, we use \eqref{expfermi} and the proof is over.

\qed

\section{Acknowledgments}

H.C. was partially supported by the Danish FNU grant {\it Mathematical
  Physics}. B.S. thanks Cyril Levy for fruitful discussions.


\begin{thebibliography}{99}
\bibitem{ABN} Angelescu N., Bundaru M., Nenciu G.: On the Landau
    Diamagnetism. Commun. Math. Phys. {\bf 42}, 9-28 (1975)

\bibitem{Adams} Adams E.N.: Magnetic susceptibility of a diamagnetic electron gas - The role of small effective electron mass. Phy. Rev. {\bf 89} (3),
633-647 (1953)

\bibitem{BC} Briet, P., Cornean, H.D.: Locating the spectrum for
 magnetic Schr\"odinger and Dirac operators. Comm. Partial
 Differential Equations {\bf 27} (5-6), 1079-1101 (2002)

\bibitem {BrCoLo1} {Briet P., Cornean H.D., Louis D.:} Diamagnetic expansions for perfect quantum gases.  J. Math. Phys.  {\bf 47} (8) 083511, (2006)

\bibitem{BrCoLo2} Briet P., Cornean H.D., Louis D.:  Generalized susceptibilities
 for a perfect quantum gas. Markov Process. Related Fields, {\bf
 11}, 177-188 (2005)

\bibitem{BrCoLo3} Briet P., Cornean H.D., Louis D.:
    Diamagnetic expansions for perfect quantum gases II: uniform
    bounds. Asymptotic Analysis  {\bf 59} (1-2), 109-123 (2008)

\bibitem{BCS1} Briet P., Cornean H.D., Savoie B.: Diamagnetism
    of quantum gases with singular potentials, J. Phys. A: Math. Theor. {\bf 43},
474008 (2010)

\bibitem{BCZ} Briet P., Cornean H.D., Zagrebnov V.: Do Bosons Condense in a Homogeneous Magnetic Field ?, J. Stat. Phys. {\bf 116},  1545-1578 (2004)
\bibitem{BL} Blount E.I.: Bloch Electrons in a Magnetic Field. Phys. Rev. {\bf 126},
1636-1653 (1962)
\bibitem{BS} Berezin F.A., Shubin M.A.: The Schr\"odinger
    Equation, Kluwer Academic Publishers, Dordrecht, 1991

\bibitem{CR} Combescure M., Robert D.: Rigorous semiclassical results for the
magnetic response of an electron gas. Rev. Math. Phys. {\bf 13} (9), 1055-1073 (2001)

\bibitem{Cor1} Cornean H.D.:
On the magnetization of a charged Bose gas in the canonical ensemble.
 Commun. Math. Phys. {\bf 212} (1), 1-27 (2000)

\bibitem{CN0} Cornean H.D., Nenciu G.: On eigenfunction decay for two dimensional magnetic Schr\"odinger operators. Commun. Math. Phys. {\bf 198} (3), 671-685 (1998)

\bibitem{CNP} Cornean H.D., Nenciu G., Pedersen T.G.:
The Faraday effect revisited: general theory. J. Math. Phys.  {\bf 47} (1), 013511 (2006)

\bibitem{CN3} Cornean H.D., Nenciu G.: The Faraday effect revisited: Thermodynamic limit.
 J. Funct. Anal. {\bf 257} (7), 2024-2066 (2009)

\bibitem{CN} Cornean H.D., Nenciu G.: Faraday effect revisited:
sum rules and convergence issues.  J. Phys. A: Math. Theor. {\bf 43}, 474012 (2010)

\bibitem{DIM} Doi S., Iwatsuka A., Mine T.:
The Uniqueness of the Integrated Density of States for the Schr\"odinger Operators with Magnetic Fields. Math. Zeit. {\bf 237}, 335-371 (2001)

\bibitem{Gla} Glasser M.L.: Magnetic properties of nearly free electrons, nonoscillatory magnetic susceptibility. Phys. Rev. {\bf 134} (5A), 1296-1299 (1964)

\bibitem{H} Huang K.: Statistical Mechanics (Second Edition). John Wiley \& Sons, 1987
\bibitem{HLS} Hebborn J.E., Luttinger J.M., Sondheimer E.H., Stiles P.J.:
The Orbital Diamagnetic Susceptibility of Bloch Electrons. J. Phys. Chem. Solids {\bf 25},
 741-749 (1964)
\bibitem{HS1} Hebborn J.E., Sondheimer E.H.:
Diamagnetism of Conduction Electrons in Metals. Phys. Rev. Letters {\bf 2},  150-152 (1959)
\bibitem{HS2} Hebborn J.E., Sondheimer E.H.: The Diamagnetism of
    Conduction Electrons in Metals. J. Phys. Chem. Solids {\bf 13},  105-123 (1960)


\bibitem{HeMo} Helffer B., Mohamed A.:
Asymptotics of the density of states for the Schr\"odinger
operator with periodic potential. Duke Math. J. {\bf 92}, 1-60 (1998)

\bibitem{HeSj}  Helffer B., Sj{\"o}strand J.:
On diamagnetism and the de Haas-Van Alphen  effect.
Annales de l'IHP, section Physique th\'eorique {\bf 52}, 303-375
(1990)
\bibitem{K} Kato T.: Perturbation Theory for Linear Operators.
Springer-Verlag, Berlin- Heidelberg, 1976
\bibitem{KK} Kjeldaas T., Kohn W.: Theory of the Diamagnetism of Bloch Electrons.
Phys. Rev. {\bf 105}, 806-813 (1957)
\bibitem{KW} Kirsch W.: Random Schr\"odinger Operators. Lect. Notes in Phys. {\bf 345},
264-370 (1989)
\bibitem{KS} Kirsch W., Simon B.:
Comparison Theorems for The Gap of Schr\"odinger Operators. J. Funct. Anal. {\bf 75}, 396-410
(1987)
\bibitem{Ku} Kuchment P.: Floquet Theory for Partial Differential Equations.
Birkhauser Verlag, Basel, 1993

\bibitem{Lan} Landau L.: Diamagnetismus der Metalle. Zeitschrift f\"ur Physik A. Hadrons and Nuclei. {\bf 64} (9-10),  629-637 (1930)


\bibitem{MK} Misra P.K., Kleinman L.:
Theory of the Diamagnetic Susceptibility of Bloch Electrons. Phys. Rev. B {\bf 5} (11),
4581-4597 (1972)
\bibitem{MR} Misra P.K., Roth L.M.: Theory of Diamagnetic Susceptibility of Metals.
Phys. Rev. {\bf 177} (3), 1089-1102  (1969)
\bibitem{N} Nenciu G.: Dynamics of band electrons in electric and magnetic fields: rigorous justification of the effective Hamiltonians. Rev. Mod. Phys. {\bf 63}, 91-127 (1991)
\bibitem{Pei} Peierls R.: Zur Theorie des Diamagnetismus von Leitungselektronen.
Zeitschrift f\"ur Physik A. Hadrons and Nuclei. {\bf 80}, 2-19 (1933)

\bibitem{Re} Resta R.: Electric polarization and orbital magnetisation: the modern theories.
J. Phys.: Condens. Matter A. Hadrons and Nuclei. {\bf 22} (11-12), 763-791 (2010)

\bibitem{Rot} Roth L.M.: Theory of Bloch Electrons in a Magnetic Field.
J. Phys. Chem. Solids {\bf 23}, 433-446 (1962)
\bibitem{RS2} Reed M., Simon B.: Methods of Modern Mathematical Physics, II : Fourier Analysis and Self-Adjointness. Academic Press, San Diego, 1975

\bibitem{Ru} Ruelle D.: Statistical Mechanics - Rigorous Results. W.A. Benjamin, New York 1969

\bibitem{BCS2} Savoie B.: PhD thesis (2010)
\bibitem{WU} Wannier G.H., Upadhyaya U.N.: Zero-Field Susceptibility of Bloch Electrons.
Phys.Rev. {\bf 136}, A803-A810 (1964)
\end{thebibliography}
\end{document}